\documentclass[11pt, a4paper]{article}
\pdfoutput=1

\usepackage{jcappub}

\usepackage{comment}
\usepackage{amsmath}
\usepackage{amsfonts}
\usepackage{amssymb}
\usepackage{graphicx}
\usepackage{mathrsfs}
\usepackage{latexsym}
\usepackage{color}
\usepackage{slashed, cancel}
\usepackage{hyperref}
\usepackage{cleveref}
\usepackage{float}
\usepackage{tikz}
\usepackage{bbold}
\usepackage{soul} 
\usepackage{mathtools}
\usepackage{support-caption} 
\usepackage{subcaption}
\usepackage{enumitem}
\usepackage{journals}
\allowdisplaybreaks

\usepackage[utf8]{inputenc} 
\usepackage[version=3]{mhchem}
\hypersetup{urlcolor=blue, colorlinks=true} 

\usepackage{fancyhdr}
\renewcommand{\arraystretch}{1}
\numberwithin{equation}{section}

\definecolor{orange}{rgb}{1,0.4,0}
\definecolor{green}{rgb}{0,0.65,0}
\definecolor{rossos}{rgb}{0.8,0.2,0.3}
\definecolor{bluscuro}{rgb}{0.15, 0.2, .85}
\definecolor{bluchiaro}{cmyk}{1,.3,0.,0.1}
\hypersetup{colorlinks, citecolor=bluscuro, linkcolor=bluscuro, urlcolor=bluscuro}

\makeatother   
\newcommand{\GeV}{{\rm \,GeV}}
\newcommand{\TeV}{{\rm \,TeV}}
\newcommand{\MeV}{{\rm \,MeV}}
\newcommand{\keV}{{\rm \,keV}}

\newcommand{\cm}{{\rm \,cm}}

\newcommand{\km}{{\rm \,km}} 
\newcommand{\s}{{\rm \,s}}
\newcommand{\g}{{\rm \,g}}
\newcommand{\K}{{\rm \,K}}

\newcommand{\Gyr}{{\rm \,Gyr}}
\newcommand{\yrs}{{\rm \,yrs}}

\newcommand{\pc}{{\rm \,pc}}
\newcommand{\kpc}{{\rm \,kpc}}

\newcommand{\Msun}{M_\odot}
\newcommand{\Mstar}{M_\star}
\newcommand{\Rstar}{R_\star}
\newcommand{\vstar}{v_\star}
\newcommand{\tstar}{t_\star}
\newcommand{\Tstar}{T_\star}

\newcommand{\Teff}{T_{\rm eff}}

\newcommand{\vesc}{v_{esc}}

\newcommand{\muFe}{\mu_{F,e}}

 \newcommand{\qomax}{q_0^{\rm MAX}}

\newcommand{\fMB}{f_{\rm MB}}
\newcommand{\fFD}{f_{\rm FD}}

\newcommand{\sigmath}{\sigma_{th}}

\newcommand{\erf}{{\rm \,Erf}}
\newcommand{\optdepth}{\tau_{\chi}}

\newcommand{\MsqT}{|\overline{M_T}|^2}

 \def\be   {\begin{equation}}   \def\ee   {\end{equation}}
 \def\ba   {\begin{array}}      \def\ea   {\end{array}}
 \def\bea  {\begin{eqnarray}}   \def\eea  {\end{eqnarray}}
 \def\bean {\begin{eqnarray*}}  \def\eean {\end{eqnarray*}}
 \def\nn{\nonumber}

\begin{document}

\hfill IPPP/20/54

\title{Improved Treatment of Dark Matter Capture in White Dwarfs}

\author[a]{Nicole F.\ Bell,}
\author[b]{Giorgio Busoni,}
\author[c]{Maura E. Ramirez-Quezada,}
\author[a]{\\Sandra Robles}
\author[a]{and Michael Virgato}
\affiliation[a]{ARC Centre of Excellence for Dark Matter Particle Physics, \\
School of Physics, The University of Melbourne, Victoria 3010, Australia}
\affiliation[b]{Max-Planck-Institut fur Kernphysik, Saupfercheckweg 1, 69117 Heidelberg, Germany}
\affiliation[c]{Institute for Particle Physics Phenomenology, Department of Physics, \\
Durham University, South Road, Durham DH1 3LE, United Kingdom}

\emailAdd{n.bell@unimelb.edu.au}
\emailAdd{giorgio.busoni@mpi-hd.mpg.de}
\emailAdd{maura.e.ramirez-quezada@durham.ac.uk}
\emailAdd{sandra.robles@unimelb.edu.au}
\emailAdd{mvirgato@student.unimelb.edu.au}

\abstract{
White dwarfs, the most abundant stellar remnants, provide a promising means of probing dark matter (DM) interactions, complimentary to terrestrial searches. The scattering of dark matter from stellar constituents leads to gravitational capture, with important observational consequences.  In particular, white dwarf heating occurs due to the energy transfer in the dark matter capture and thermalisation processes, and the subsequent annihilation of captured dark matter. We consider the capture of dark matter by scattering on either the ion or the degenerate electron component of white dwarfs. For ions, we account for the stellar structure, the star opacity, realistic nuclear form factors that go beyond the simple Helm approach, and finite temperature effects pertinent to sub-GeV dark matter. Electrons are treated as relativistic, degenerate targets, with Pauli blocking, finite temperature and multiple scattering effects all taken into account. We also estimate the dark matter evaporation rate. The dark matter-nucleon/electron scattering cross sections can be constrained by comparing the heating rate due to dark matter capture with observations of cold white dwarfs in dark matter-rich environments. We apply this technique to  observations of old white dwarfs in the globular cluster Messier~4, which we assume to be located in a DM subhalo. For dark matter-nucleon scattering, we find that white dwarfs can probe the sub-GeV mass range inaccessible to direct detection searches, with the low mass reach limited only by either  evaporation or dominant DM annihilation to neutrinos, and can be competitive with direct detection in the $1-10^4\GeV$ range. White dwarf limits on dark matter-electron scattering are found to outperform current electron recoil experiments over the full mass range considered, and extend well beyond the $\sim 10\GeV$ mass regime where the sensitivity of electron recoil experiments is reduced.}

\maketitle

\section{Introduction}

Despite significant improvement in the sensitivity of dark matter (DM) direct detection (DD) experiments in recent years, the nature of dark matter remains an open question in modern physics. As these experiments are limited in their reach by practical detector masses, detectable target recoils and irreducible backgrounds (such as the neutrino floor), alternative means of observing DM are highly desirable. One such avenue is the possibility that DM interactions with astrophysical objects can lead to detectable signals. The capture of DM in stars  has long been studied in this context~\cite{Gould:1987ir, Gould:1987ju, Goldman:1989nd, Jungman:1995df,Kouvaris:2007ay, Bertone:2007ae,Kouvaris:2010vv,deLavallaz:2010wp, Busoni:2013kaa,Garani:2017jcj,Busoni:2017mhe,Ilie:2020nzp,Ilie:2020iup}, as the accumulated DM may annihilate in the stellar interior, leading to neutrino signals~\cite{Tanaka:2011uf,Choi:2015ara,Bell:2021esh,Adrian-Martinez:2016ujo,Adrian-Martinez:2016gti,Aartsen:2016zhm}  or, potentially, to cosmic rays and gamma rays~\cite{Batell:2009zp, Schuster:2009au, Bell:2011sn, Feng:2016ijc, Leane:2017vag,Cermeno:2018qgu,Bell:2021pyy}.

Recently, there has been a renewed interest in the capture of DM within neutron stars (NSs)~\cite{Bramante:2017xlb,Baryakhtar:2017dbj, Raj:2017wrv,Bell:2018pkk, Garani:2018kkd, Bell:2019pyc, Joglekar:2019vzy,Acevedo:2019agu,Joglekar:2020liw, Bell:2020jou, Ilie:2020vec, Bell:2020lmm, Bell:2020obw} and, to a lesser extent, white dwarfs (WDs)~\cite{Bertone:2007ae,McCullough:2010ai,Hooper:2010es,Amaro-Seoane:2015uny,Panotopoulos:2020kuo,Curtin:2020tkm} where the focus for the latter has been on the multiple scattering regime relevant for large DM masses~\cite{Bramante:2017xlb, Dasgupta:2019juq}. This interest has been driven by the potential heating of these objects due to the capture, thermalisation, and subsequent annihilation of dark matter. The observational effect of this heating is more pronounced for older compact  stars as, in the absence of additional heating mechanisms, they are expected to cool to low temperatures. Hence, observations of old, cold, isolated, compact objects would enable rather strict bounds to be placed on the strength of DM interactions with stellar constituents. For many interaction types, these bounds would potentially exceed those arising from current direct detection experiments, in some cases by many orders of magnitude.  Indeed, there is potential to obtain strong constraints for a huge range of DM masses, extending from very low DM mass, where any nuclear or electron recoil signal would lie well below experimental detection thresholds, up to very DM large mass, where direct detection experiments lose sensitivity.

In this paper, we study DM capture in WDs and the resulting heating. The advantage of considering WDs, rather than NSs, is due to the existence of observational data such as the recent releases from the Gaia mission~\cite{GaiaDR2:2018, GaiaDR3:2020, GentileFusillo:2019, Tremblay:2020, McCleery:2020}. As such, the physics of WDs is much better constrained than that of NSs, with a well defined mass-radius relation (and hence much less uncertainty in the equation of state), as well as luminosity-age relations which are related to the spectral evolution of these stars. In the absence of anomalous cooling of old, isolated  WDs, we can equate the observed luminosity of WDs that lie in DM-rich environments with the heating rate due to DM annihilation.
This allows us to place limits on the DM interaction strength with WD constituents. 
Such observational data have been collected for WDs in the globular cluster Messier~4 (M4) using the Hubble Space Telescope~\cite{Bedin:2009}. We use  these observations to derive limits, assuming that M4 was formed in a DM sub-halo~\cite{McCullough:2010ai}.

The WD core is composed of an ionic lattice and a degenerate electron gas.  We shall consider the scattering of DM with either of these components, and hence will derive limits on both DM-nucleon and DM-electron scattering cross sections.  
In order to calculate the extent to which WDs are heated due to DM capture and annihilation, we first calculate the rate at which DM is captured by the stars, accounting for scattering from either ions or electrons. The former can be treated using methods relevant for non-relativistic targets~\cite{Busoni:2017mhe, Garani:2017jcj}, while the highly degenerate electrons require the use of relativistic kinematics, as well as a proper treatment of Pauli blocking~\cite{Bell:2020jou, Bell:2020lmm}. For ion targets, we restrict our attention to DM masses $\lesssim 1$ PeV, avoiding the region where multiple scattering dominates the capture process.  For electron targets, we consider the same DM mass range while, accounting for multiple scattering effects. (Multiple scattering becomes relevant for lower DM mass for the case of scattering on electrons compared to ions.)
We then consider DM masses down to the evaporation mass for the relevant WD and target. This is the smallest mass for which the captured DM is not efficiently evaporated from the star due to up-scattering.

The paper is then organised as follows: We describe the structure and observations of WDs in section~\ref{sec:wd}. The details of capture due to scattering from ions are discussed in section~\ref{sec:capions}, including the optically thin limit, the treatment of WD opacity, finite temperature effects and evaporation. Section~\ref{sec:capelectrons} then details the capture process for the case of scattering from electron targets, both for single and multiple scattering, as well as the effects due to finite temperature. We present our results in section~\ref{sec:results} and concluding remarks are given in section~\ref{sec:conclusions}.

\section{White Dwarfs}
\label{sec:wd}
The fate of main sequence stars of mass below $\Mstar\lesssim 8 -10 \, \Msun$ is to end their life cycles as white dwarfs. Consequently, these compact stellar remnants, which are supported against gravitational collapse by electron degeneracy pressure, are the most abundant stars in the Galaxy ($\gtrsim90\%$).  They are born at very high temperatures and cool down over billions of years. Observations of the coldest WDs therefore contain information on the star formation history of the Galaxy. We outline the internal structure of WDs below, and summarise the current observational status.

\subsection{Internal structure and Equation of State}
\label{sec:WDmodels}

The vast majority of observed WDs are composed primarily of carbon and oxygen, plus small traces of elements heavier than helium. 
At the extremely high densities found in WDs $\sim10^6-10^{10}\g\cm^{-3}$, electrons are strongly degenerate and determine the WD equation of state (EoS) and internal structure.  The stellar core resembles a Coulomb lattice of ions surrounded by degenerate electrons, which implies that the WD core is isothermal and a very good thermal conductor. 
The degenerate core is enclosed by a thin envelope that  accounts for $\lesssim1\%$  of the total mass~\cite{Fontaine:2001}. 
The outer layers form an atmosphere which is rich in lighter elements such as  hydrogen or helium, where the exact composition depends on the evolution of the WD progenitor and changes as the WD cools. 
This atmosphere is non-degenerate and extremely opaque to radiation, with an EoS that is subject to finite temperature effects. 

In the limit of zero temperature, the simplest way to obtain the WD EoS is to assume an ideal Fermi gas of degenerate electrons, for a WD that is primarily composed of a single element. Corrections to the non-interacting electron picture were introduced early by Salpeter~\cite{Salpeter:1961}. By introducing the Wigner-Seitz (WS) cell approximation and assuming point-like nuclei, Salpeter obtained an analytical EoS that accounts  for interactions between electrons and ions as well as other Coulomb corrections. These corrections, in general, depend on the chemical composition of the star. 

More recently, it has been shown that the treatment of matter at high pressures presented by Feynman, Metropolis and Teller~\cite{Feynman:1949zz} can be extended to consistently take into account weak interactions and relativistic effects \cite{Rotondo:2009cr, Rotondo:2011zz}, and incorporates Coulomb corrections in a more natural manner than the Salpeter EoS. The resulting Feynman-Metropolis-Teller (FMT) EoS is obtained by considering a relativistic Thomas-Fermi model within Wigner-Seitz cells of radius $R_{WS}$. 
For degenerate, relativistic, electrons, the equilibrium condition is that the Fermi energy, $E_e^F$, is constant within the cell,
\begin{equation}\label{eq:rel_equil}
E_e^F=\sqrt{(p_e^F)^2+m_e^2}-m_e-eV(r) = \rm{constant},
\end{equation}
where $V(r)$ is the Coulomb potential inside the cell, $p_e^F$ is the electron Fermi momentum, $m_e$ is the electron mass and $e$ is the electric charge. In order to obtain an integrable solution for the energy density near the origin, it is necessary to introduce a finite size for the nucleus, with radius $ R_c = \Delta\lambda_{\pi} Z^{1/3}$, 
where $\lambda_\pi$ is the pion Compton wavelength, $\Delta \approx (r_0 /\lambda_\pi)(A/Z)^{1/3}$, $Z$ is the proton number, $A$ is the atomic mass, and $r_0$ is an empirical constant $\sim 1.2\;\text{fm}$. The proton and electron number densities inside the cell are then given by
\begin{align}
    n_p &=  \frac{(p^F_p)^3}{3\pi^2} =\frac{3Z}{4\pi R_c^3}\theta( R_c - r ) = \frac{3}{4\pi} \left( \frac{1}{\Delta \lambda_\pi} \right)^3 \theta(R_c -r), \label{eq:prot_dens}\\
    n_e &= \frac{(p^F_e)^3}{3\pi^2} = \frac{1}{3\pi^2}\left[ \hat{V}^2(r) + 2m_e \hat{V}(r)\right]^{3/2},\label{eq:elec_dens}
\end{align}
where 
\begin{equation}
    \hat{V}(r) = eV(r) + E_e^F. \label{eq:vhat}
\end{equation}
Using these expressions in the Poisson equation for the Coulomb potential results in the relativistic Thomas-Fermi equation
\begin{equation}
    \frac{1}{3x}\frac{d^2\beta}{dx^2} = -\frac{\alpha}{\Delta^3}\theta(x_c -x) + \frac{4\alpha}{9\pi}\left[ \frac{\beta^2(x)}{x^2} + 2\frac{m_e}{m_\pi}\frac{\beta(x)}{x}\right]^{3/2},\label{eq:FMT_DE}
\end{equation}
which is written in terms of the dimensionless quantities $x = r/\lambda_\pi$ and $\beta(r) = r\hat{V}(r)$, such that $x_c = R_c/\lambda_\pi$ and $x_{WS} = R_{WS}/\lambda_\pi$, and $\alpha$ is the fine structure constant.
The requirement of global charge neutrality imposes the conditions $dV/dr |_{r=R_{WS}} =0$ and $V(R_{WS}) = 0$, which translate to the boundary conditions
\begin{equation}
  \beta(0) = 0, \qquad
  \left. \frac{d\beta}{dx}\right|_{x_{WS}} = \frac{\beta(x_{WS})}{x_{WS}}. \label{eq:TF_bc}
\end{equation}
By solving these equations, we are able to obtain the relevant thermodynamic quantities, namely the electron and proton  number densities, electron chemical potential, and the energy and pressure of the cell. The electron chemical potential is obtained by evaluating Eq.~\ref{eq:rel_equil} at the cell radius, noting that the Coulomb potential must vanish there, which results in the usual expression
\begin{equation}
    \mu_{F,e} = \sqrt{(p_e^F)^2+m_e^2}-m_e.\label{eq:mufe}
\end{equation}
The energy and pressure of the cell can then be obtained following the analysis presented in~ref.~\cite{Rotondo:2011zz}. The cell energy gains contributions from the nuclear mass, electron kinetic energy and Coulomb interactions, such that
\begin{align}
    E_{tot} & = M_N + E_k + E_C,\\
    E_k & = \int_0^{R_{WS}}4\pi r^2 (\mathcal{E}_e(r) - m_e n_e(r))\;dr,\\
    E_C & = \frac{1}{2}\int_{R_c}^{R_{WS}}4\pi r^2 e(n_p(r) - n_e(r))V(r)\;dr,
\end{align}
where 
\begin{equation}
    \mathcal{E}_e(r) = \frac{1}{\pi^2}\int_0^{p_e^F}p^2\sqrt{p^2 + m_e^2}\;dp,
\end{equation}
is the electron energy density, and  $M_N$ the mass of the nucleus. The only contribution to the internal cell pressure comes from the electrons,
\begin{equation}
    P_e(r) = \frac{1}{3\pi^2}\int_0^{p_e^F}\frac{p^4}{\sqrt{p^2+m_e^2}}\;dp,
\end{equation}
with the total pressure of the cell being $P_{tot} = P_e(R_{WS})$.
Finally, the EoS is then obtained by solving Eq.~\ref{eq:FMT_DE} for various cell radii, yielding a relation between $E_{tot}$ and $P_{tot}$ paramaterised by the radius of the Wigner-Seitz cell.

\begin{table}[t]
  \centering
    \begin{tabular}{|l|l|l|l|l|} 
    \hline
      \textbf{EoS} & \textbf{WD}$_\mathbf{1}$  & \textbf{WD}$_\mathbf{2}$&\textbf{WD}$_\mathbf{3}$ &\textbf{WD}$_\mathbf{4}$\\
      \hline
     $\rho_c\,[\g\cm^{-3}]$ &  $1.98\times10^{6}$ & $3.46\times10^{7} $ & $3.02\times 10^{8}$ & $1.14\times10^{10}$\\
     $M_\star\,[M_\odot]$ & $0.490$ &  $1.000 $ & $1.252$ &$1.384$\\
     $R_\star\,[\km]$ & $9.39\times10^{3}$ &  $5.38\times10^{3}$ & $3.29\times 10^3$ & $1.25\times10^3$\\
     $\vesc(\Rstar) \, [\km/\s]$ & $3.72\times10^{3}$ & $7.03\times10^{3}$ & $1.01\times10^{4}$ & $1.71\times10^{4}$ \\
      \hline
    \end{tabular}
  \caption{Four configurations for white dwarfs composed of carbon, with a FMT EoS, where $\rho_c$ is the  central density. }
    \label{tab:WDs}
\end{table}

\begin{figure}[t]
    \centering
    \includegraphics[width=0.5\textwidth]{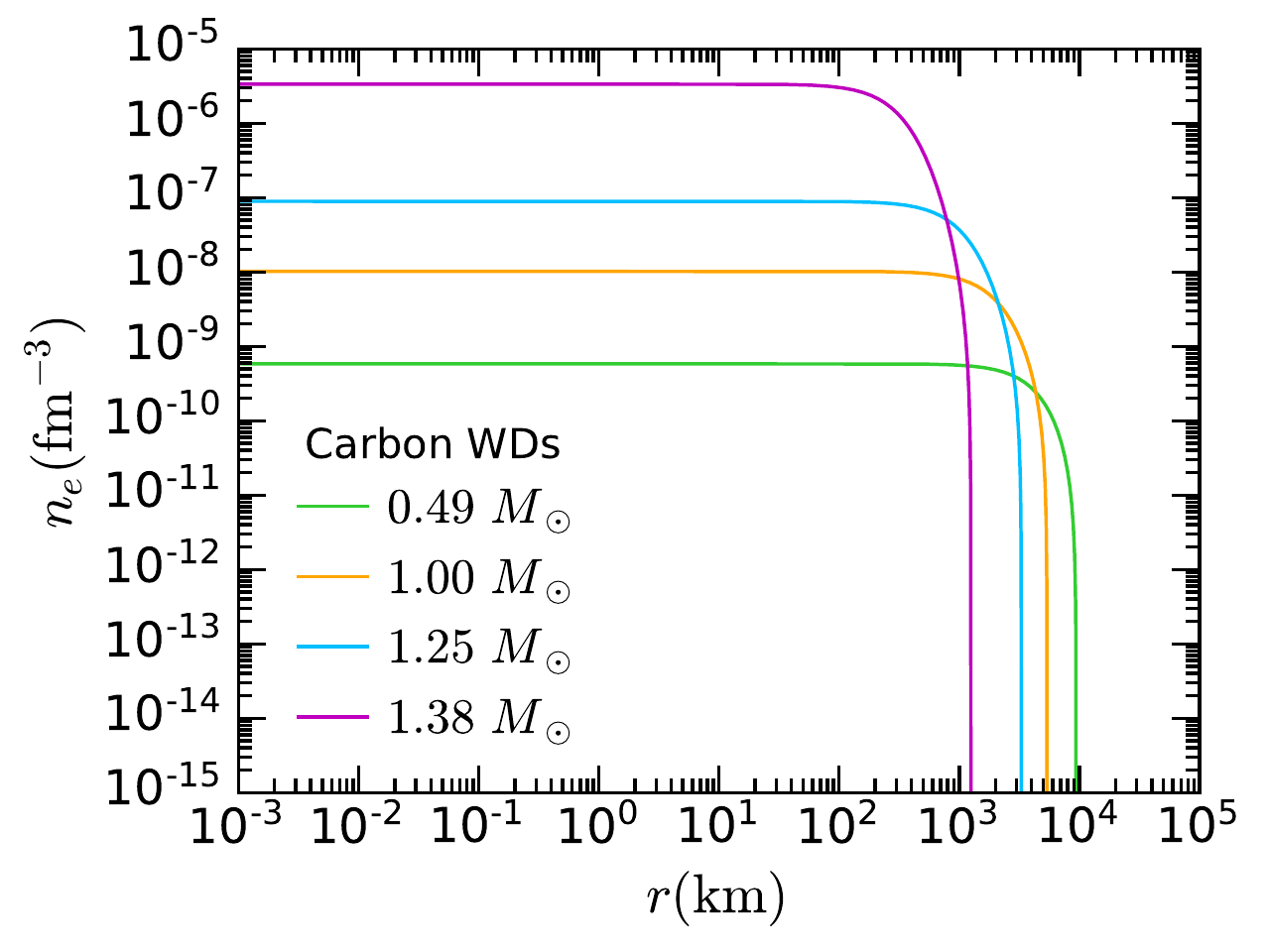}
    \includegraphics[width=0.49\textwidth]{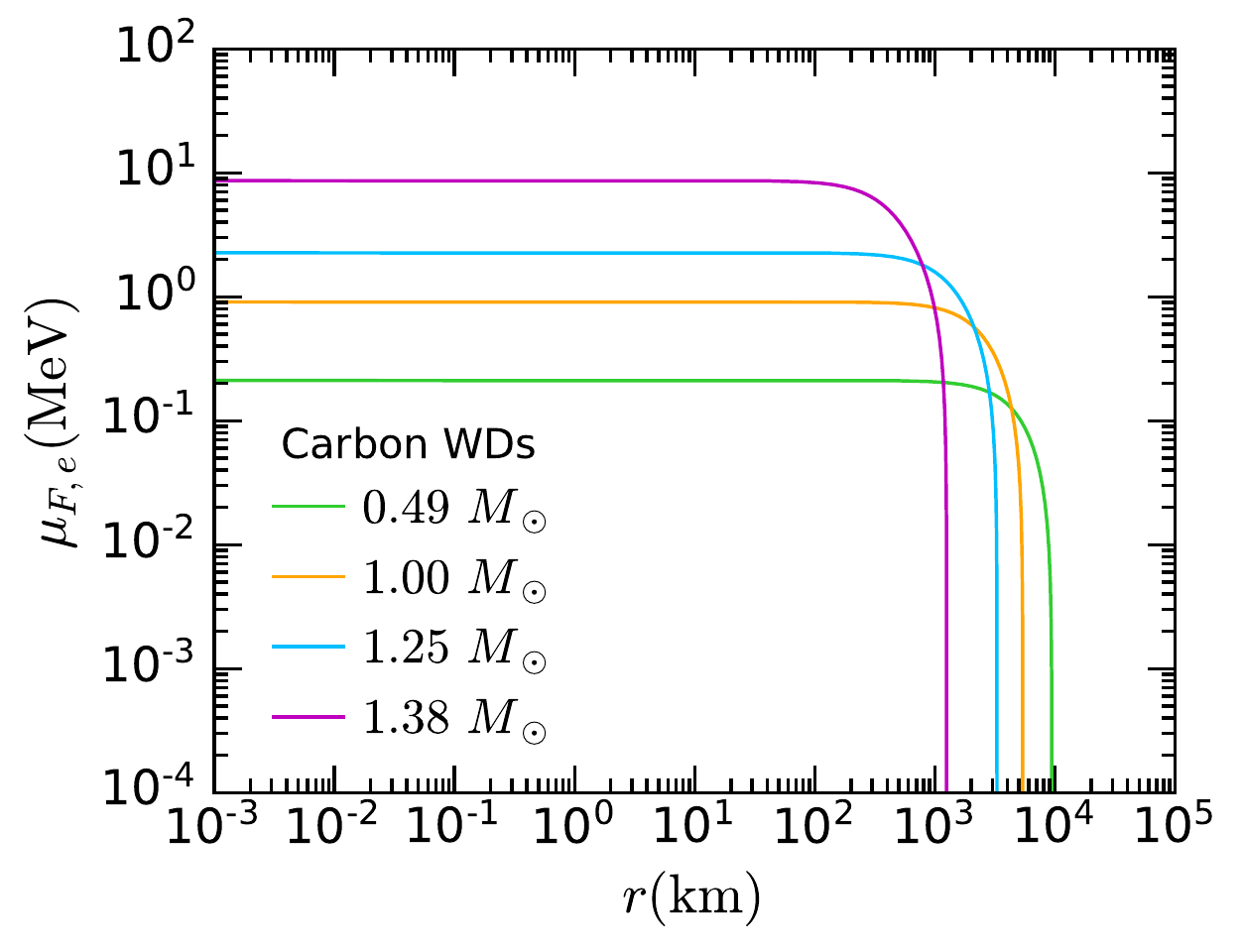}    
    \includegraphics[width=0.5\textwidth]{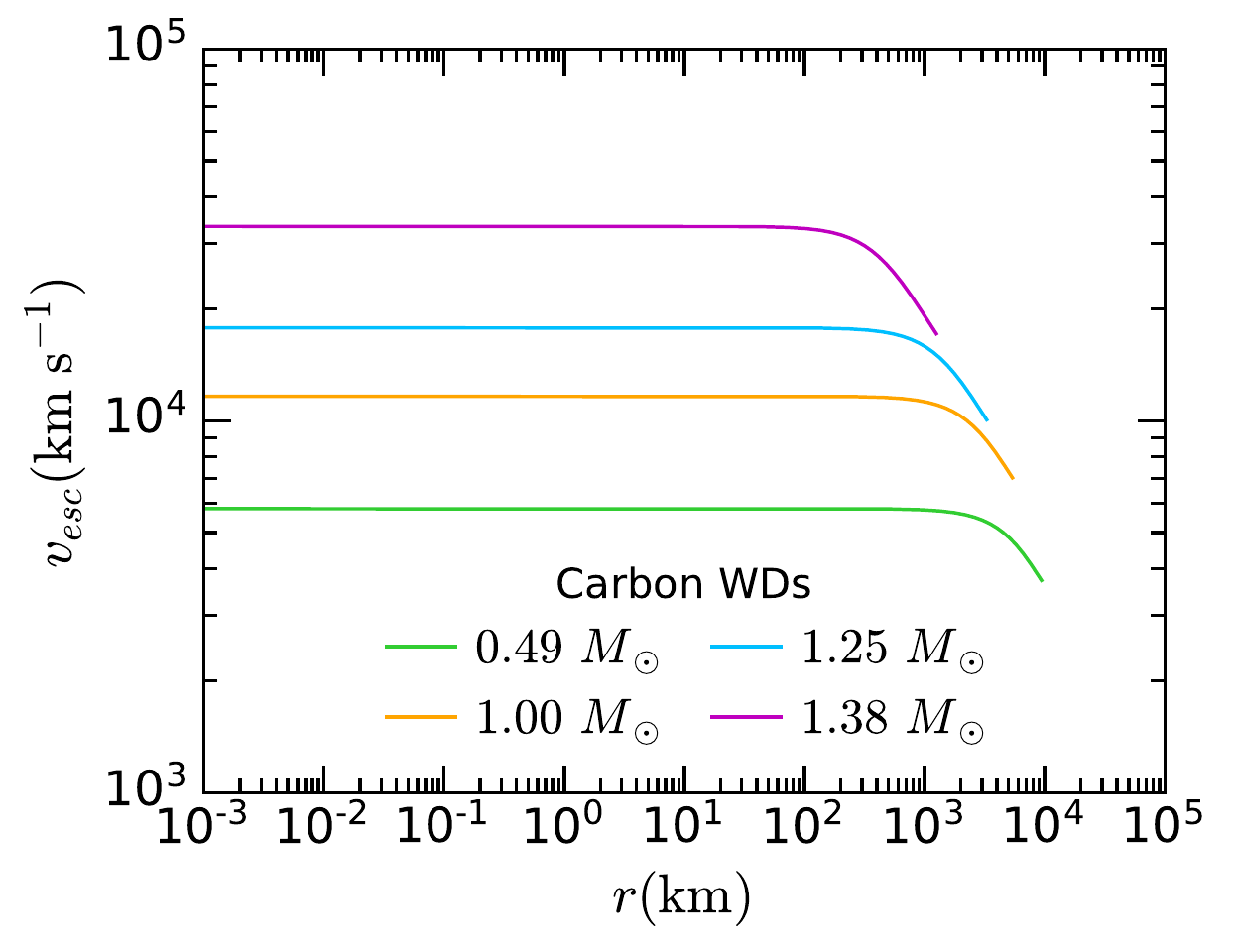}    
    \caption{Electron number density (top left), chemical potential (top right) and escape velocity (bottom) radial profiles for the carbon WDs with FMT EoS in Table~\ref{tab:WDs}. }
    \label{fig:WDradprofs}
\end{figure}

Different WD configurations can be obtained, assuming a non-rotating spherically symmetric star, by solving the
Tolman-Oppenheimer-Volkoff (TOV) equations~\cite{Tolman:1939jz,Oppenheimer:1939ne} coupled to the FMT EoS with different initial conditions for the pressure at the centre of the star. In Fig.~\ref{fig:WDradprofs} we show radial profiles for $n_e$ (top left), $\muFe$ (top right) and escape velocity $\vesc$ (bottom) for the carbon WDs in Table~\ref{tab:WDs}. Note that the difference in radius between the lightest and heaviest WD in Table~\ref{tab:WDs} spans almost one order of magnitude, while the electron number densities in the core can vary up to 4 orders of magnitude (see top left panel). As expected, electrons are more degenerate in more compact WDs and become relativistic (see top right panel). The escape velocity can reach ${\cal O}(0.1\,c)$ at the interior of the most compact WDs, while for very low mass WDs 
it can be as low as $\sim0.003\,c$.

The mass-radius relations obtained from a zero-temperature EoS begin to deviate from observations for low mass WDs. To address this discrepancy, finite temperature effects can be introduced to the EoS~\cite{deCarvalho:2013rea}. When calculating the effects for carbon WDs with core temperatures $\Tstar=10^6-10^7$K, we find that the chemical potential is smaller by only $\sim16\%$, and the number density by $\sim~35\%$.
As we shall focus on old WDs, we shall assume $\Tstar=10^5\K$, for which the zero-temperature regime holds. Note that finite temperature effects on the EoS may be more pronounced for light WDs composed of helium~\cite{deCarvalho:2013rea}.

\subsection{Observations}
\label{sec:WDobserv}

The rate at which the energy of the WD core is radiated away is determined by the outer non-degenerate layers. 
Spectroscopic observations shed light on  the  composition of these layers, and can be used to classify WDs in terms of $\sim$ six spectral types. 
Most of the observed WDs lie in the DA (hydrogen-rich) and DB (helium-rich) categories.  Note that as WDs slowly cool, they undergo spectral evolution. There is a well defined relation between their luminosity and age (cooling time) that, together with recent breakthroughs in theory and observations, allow to estimate the age of the stars in the solar neighbourhood and to date the nearest star  clusters~\cite{Hansen:2004ih,Hansen:2007ve,Bedin:2009,Hansen:2013,Kilic:2017}. 

Over the past few decades, WDs have been extensively observed using photometry and spectroscopy. Most of the WDs, spectroscopically confirmed,  have been discovered by large area surveys, such as the Sloan Digital Sky Survey (SDSS)~\cite{York:2000gk}. However, these local samples are dominated by young WDs with relatively high effective temperatures\footnote{The effective  temperature is the temperature that characterises the surface of the star, assuming that WDs are perfect blackbody emitters, i.e.  $L_\gamma = 4\pi \sigma_{SB} \, \Rstar^2 \, \Teff^4$, where $\sigma_{SB}$ is the Stefan–Boltzmann constant.} ($\Teff\gtrsim10^4\K$)~\cite{Eisenstein:2006ty,Kleinman:2013,Tremblay:2016,Kepler:2017,Kepler:2019}. 
Recently, the local volume sample of nearby stars within $\sim100\pc$ have been catalogued by the Gaia spacecraft~\cite{GaiaDR2:2018,GaiaDR3:2020}, an astrometric mission. New WD candidates have been identified~\cite{GentileFusillo:2019}, followed by dedicated spectroscopic observations~\cite{Tremblay:2020,McCleery:2020}, increasing the local sample of cool WDs ($\Teff\lesssim5000\K$). 

On the other hand, globular clusters (GCs) are the oldest known stellar systems in the Galaxy. Among them is Messier 4 (M4), also classified as NGC 6121, which is the closest globular cluster to Earth, $\sim1.9\kpc$ away~\cite{Neeley:2015,Watkins:2017,Shao:2019}. The age of M4, $11.6\Gyr$, has been estimated using observations of faint cold WDs with the Hubble Space Telescope  (HST)~\cite{Hansen:2004ih,Bedin:2009}. 
This HST data, corrected for reddening and extinction, 
was converted into luminosities and effective temperatures in ref.~\cite{McCullough:2010ai}. From these calculations, it is possible to infer WD radii and their corresponding masses assuming a mass-radius relation. 

We should note that there is no evidence so far that globular clusters are embedded in DM halos. Furthermore, there is no consensus on the scenario that leads to globular cluster formation. Observations and simulations suggest that there is more than one plausible mechanism. In the hierarchical structure formation scenario of standard cosmology, old, metal poor, globular clusters are formed  within their own halos before or shortly after the onset of reionization~\cite{Peebles:1984,Bromm:2002jt,Mashchenko:2004hj,Mashchenko:2004hk,Ricotti:2016zhz}. Old globular clusters with intermediate metallicity, such as M4~\cite{Carretta:2009a,Carretta:2009b,Wang:2017}, could have been accreted from small satellite galaxies and lost a large fraction of their initial DM content due to tidal stripping by the host galaxy~\cite{Bromm:2002jt,Mashchenko:2004hk,Saitoh:2005tt}, though a small DM fraction could have survived in the innermost region of the cluster~\cite{Mashchenko:2004hk,Ibata:2012eq}. 
Since the cool WDs observed in M4 are located in the dense core of the globular cluster, well within the tidal radius, it is expected that the DM content in this region has survived tidal disruption. Under this assumption, it is possible to estimate the DM density as outlined in refs.~\cite{Bertone:2007ae,McCullough:2010ai}.  The DM density at the largest radius the WDs were observed ($2.3 \pc$) was conservatively estimated to be $\rho_\chi =798~(531.5) \GeV\cm^{-3}$  for a contracted (uncontracted) NFW profile \cite{McCullough:2010ai}. 
It is worth noting that the DM density in M4 was found to be of the order of few $\GeV/\cm^3$ in ref.~\cite{Hooper:2010es}. However, the estimation in ref.~\cite{McCullough:2010ai} for M4, adopted in this work, is in good agreement with that of ref.~\cite{Amaro-Seoane:2015uny} for the GC Omega Centauri (NGC 5139).

\section{Capture by Scattering on Ions}
\label{sec:capions}

As discussed in section~\ref{sec:WDmodels}, there are two possible kinds of targets with which DM particles can scatter: 
the non-degenerate ions and the degenerate electrons. In this section, we shall compute the DM capture rate due to scattering on ions, assuming WDs are made of only one element.

The gravitational field of a WD is sufficiently weak that Newtonian gravity and non-relativistic kinematics are valid approximations to be used in the capture process.
Therefore, we adopt a methodology very similar to that of ref.~\cite{Busoni:2017mhe}, which is based on refs.~\cite{Gould:1987ir,Gould:1987ju,Gould:1987}.
We assume a Maxwell Boltzmann (MB) distribution, $\fMB(u_\chi)$, for DM velocity in the globular cluster M4 or in the Galactic halo, resulting in the following relative speed distribution with respect to the ions in the WD~\cite{Busoni:2017mhe}
\begin{equation}
    \fMB(u_\chi)du_\chi=\frac{u_\chi}{v_\star}\sqrt{\frac{3}{2\pi(v_d^2+3T_\star/m_T)}}
\left(\exp\left[-\frac{3\left(u_\chi-v_\star\right)^2}{2(v_d^2+3T_\star/m_T)}\right]-\exp\left[-\frac{3\left(u_\chi+v_\star\right)^2}{2(v_d^2+3T_\star/m_T)}\right]\right)du_\chi, \label{eq:veldistfiniteT}
\end{equation} 
where $u_\chi$ is the DM velocity far away from the star, $m_T$ is the mass of the ion target,  $v_d$ is the velocity dispersion of the halo, $\vstar$ is the WD velocity in the GC/Galactic rest frame and $\Tstar$ is the WD core temperature. Note that this expression holds for GC escape velocities that satisfy $v_{esc}^{GC}\gg v_\star+v_d+\sqrt{3\Tstar/m_T}$. 
In the $T_\star\rightarrow0$ limit this expression reduces to 
\begin{equation}
\lim_{T_\star\to 0}\fMB(u_\chi)du_\chi=\frac{u_\chi}{v_dv_\star}\sqrt{\frac{3}{2\pi}}
\left(\exp\left[-\frac{3}{2v_d^2}\left(u_\chi-v_\star\right)^2\right]-\exp\left[-\frac{3}{2v_d^2}\left(u_\chi+v_\star\right)^2\right]\right)du_\chi. 
\label{eq:distribution_relative_distribution}
\end{equation}

\medskip
\subsection{Optically thin limit}
\label{sec:ionthin}

Assuming that the WD is optically thin to DM scattering and that only one collision is required for DM particles to become gravitationally bound to the star, the capture rate is given by~\cite{Busoni:2017mhe}
\begin{eqnarray}
     C_{opt.\,thin} &=& \frac{\rho_\chi}{m_\chi}\int_0^{R_\star} 4\pi r^2
     \int_0^\infty du_\chi \frac{w(r)}{u_\chi} \fMB(u_\chi) \Omega^-(w),\label{eq:ioncapdef}\\
     \Omega^-(w) &=& \int_0^{v_e}dv R^-(w\to v),\label{eq:iongammadef}\\
 R^-(w\to v) &=&\int_0^\infty ds'\int_0^\infty dt'\frac{32 \mu_+^4}{\sqrt{\pi}}\kappa^3n_T(r)\frac{d\sigma_{T\chi}}{d\cos\theta}\frac{vt'}{w}e^{-\kappa^2v_T^2}\Theta(t'+s'-w)\Theta(v-|t'-s'|),
 \label{eq:ionRdef}
\end{eqnarray}
where $\rho_\chi$ is the local DM density, $m_\chi$ is the DM mass, 
$n_T$ is the  number density of the target nucleus (obtained in Section~\ref{sec:WDmodels}),  $m_T$ is  the mass of the nucleus target,  $\sigma_{T\chi}$ is the DM-ion cross section in the centre of mass (CoM) frame, 
\begin{eqnarray}
    \mu &=& \frac{m_\chi}{m_T},\qquad 
    \mu_\pm = \frac{\mu\pm 1}{2}, \qquad \kappa^2=\frac{m_T}{2\Tstar},\\
    v_T^2 &=& 2\mu\mu_+ (t')^2+2\mu_+ (s')^2-\mu w^2,
\end{eqnarray} 
$s'$ and $t'$ are the velocity of the CoM and initial DM velocity in the CoM frame respectively,
and $w$ is the speed of DM at a finite distance from the centre of the star, given by 
\begin{equation}
    w^2(r)=u_\chi^2+v_{esc}(r)^2,
\end{equation}
where $v_{esc}(r)$ is the escape velocity at a distance $r$ from the centre of the star.

An absolute upper limit on the capture rate arises when we assume the maximum capture probability  in Eq.~\ref{eq:ioncapdef}, where every DM particle traversing the star is captured. This is the so-called
geometric limit, which is given by
 \begin{equation}
 C_{geom}=\frac{\pi R_\star^2 \rho_\chi}{m_\chi} \int_0^\infty \frac{w^2(\Rstar)}{u_\chi} \fMB(u_\chi)du_\chi,\label{eq:geom_Tto0}
\end{equation}
where, in the $T_\star\rightarrow0$ limit, we find
\begin{equation}
 C_{geom}=\frac{\pi R_\star^2 \rho_\chi}{3v_\star m_\chi} \left[(3 v_{esc}^2(\Rstar)+3 v_\star^2+v_d^2) \erf \left(\sqrt{\frac{3}{2}}\frac{v_\star}{v_d}\right) +\sqrt{\frac{6}{\pi}} \vstar v_d e^{-\frac{3\vstar^2}{2 v_d^2}}\right], \label{eq:CWD_geo_limit}
\end{equation}
For finite temperature, the geometric limit is obtained by making the following replacement in the previous expression
\begin{equation}
    v_d\rightarrow\sqrt{v_d^2+3T_\star/m_T} \label{eq:vdfinT}.
\end{equation}
When $v_{esc}\gg v_\star,v_d$, the geometric limit can be approximated by
\begin{equation}
 C_{geom}\sim\frac{\pi R_\star^2 \rho_\chi}{v_\star m_\chi}v_{esc}^2(\Rstar)\erf \left(\sqrt{\frac{3}{2}}\frac{v_\star}{v_d}\right). \label{eq:CWD_geo_limit_simple}
\end{equation}

\begin{figure}
 \centering
  \includegraphics[width=0.55\textwidth]{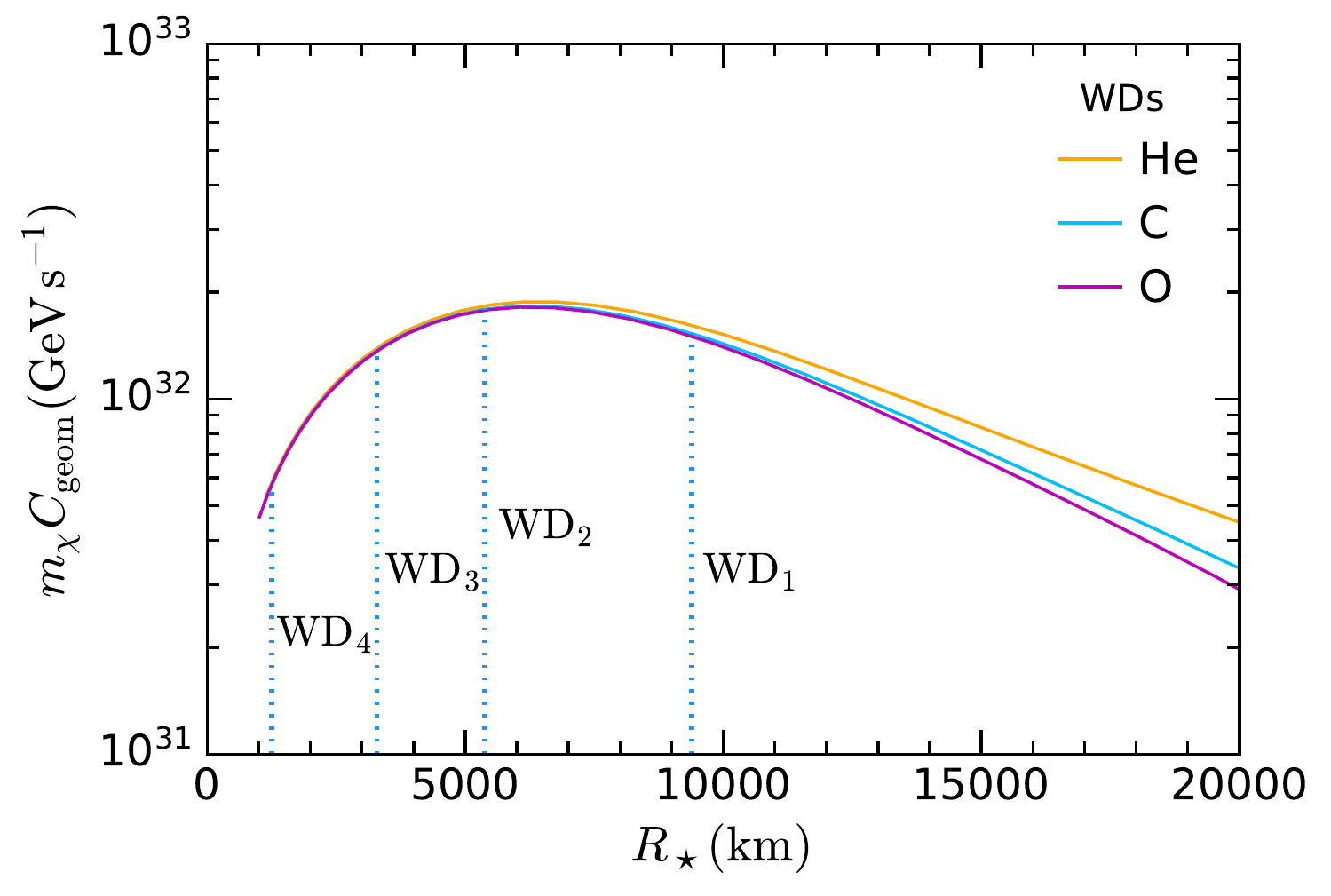}
    \caption{Capture rate in the geometric limit for different WD configurations with a FMT EoS, as a function of the stellar radius $R_\star$. The dotted blue lines represent the carbon WD configurations in Table~\ref{tab:WDs}.  }
  \label{fig:Cgeom}
\end{figure}

In the geometric limit, the DM capture rate is proportional to $1/m_\chi$  and it is independent of the DM-nucleon cross section.  In Fig.~\ref{fig:Cgeom}, we show the geometric limit in the zero temperature approximation for  WDs composed solely of He, C or O in the globular cluster M4. We have assumed the conservative values $\vstar=20\km\s^{-1}$, $v_d=8\km\s^{-1}$ and $\rho_\chi=798\GeV\cm^{-3}$ for WDs in M4, as derived in ref.~\cite{McCullough:2010ai}. As we can see, there is a maximum capture rate achievable by He, C and O WDs, which for a degenerate carbon core corresponds to a configuration with $\Mstar \sim0.83\,\mathrm{M}_\odot$ and radius $\Rstar\sim 6600\km$.  
This maximum value arises from the interplay between the mass-radius relation and the higher escape velocities of more compact WDs.

\begin{table}[t]
\centering
{\renewcommand{\arraystretch}{1.3}
\begin{tabular}{ | c | c | c | c |}
  \hline                        
  Name & Operator & Coupling & $|\overline{M}|^2(s,t)$   \\   \hline
  D1 & $\bar\chi  \chi\;\bar f  f $ & ${y_f}/{\Lambda_f^2}$ & $\frac{(c_N^S)^2}{\Lambda_f^4} \frac{\left(4 m_{\chi }^2-t\right) \left(4 m_{\chi }^2-\mu ^2
   t\right)}{\mu ^2}$ \\  \hline
  D2 & $\bar\chi \gamma^5 \chi\;\bar f f $ & $i{y_f}/{\Lambda_f^2}$ & $\frac{(c_N^S)^2}{\Lambda_f^4} \frac{t \left(\mu ^2 t-4 m_{\chi }^2\right)}{\mu ^2}$ \\  \hline
  D3 & $\bar\chi \chi\;\bar f \gamma^5  f $&  $i{y_f}/{\Lambda_f^2}$ &  $\frac{(c_N^P)^2 }{\Lambda_f^4} t \left(t-4 m_{\chi }^2\right)$ \\  \hline
  D4 & $\bar\chi \gamma^5 \chi\; \bar f \gamma^5 f $ & ${y_f}/{\Lambda_f^2}$  & $\frac{(c_N^P)^2}{\Lambda_f^4} t^2$ \\  \hline
  D5 & $\bar \chi \gamma_\mu \chi\; \bar f \gamma^\mu f$ & ${1}/{\Lambda_f^2}$ &  $2\frac{(c_N^V)^2}{\Lambda_f^4} \frac{2 \left(\mu ^2+1\right)^2 m_{\chi }^4-4 \left(\mu ^2+1\right) \mu ^2 s m_{\chi }^2+\mu ^4 \left(2 s^2+2 s t+t^2\right)}{\mu^4}$ \\  \hline
  D6 & $\bar\chi \gamma_\mu \gamma^5 \chi\; \bar  f \gamma^\mu f $ & ${1}/{\Lambda_f^2}$ &  $2\frac{(c_N^V)^2}{\Lambda_f^4} \frac{2 \left(\mu ^2-1\right)^2 m_{\chi }^4-4 \mu ^2 m_{\chi }^2 \left(\mu ^2 s+s+\mu ^2 t\right)+\mu ^4 \left(2 s^2+2 s
   t+t^2\right)}{\mu^4}$  \\  \hline
  D7 & $\bar \chi \gamma_\mu  \chi\; \bar f \gamma^\mu\gamma^5  f$ & ${1}/{\Lambda_f^2}$ &  $2\frac{(c_N^A)^2}{\Lambda_f^4} \frac{2 \left(\mu ^2-1\right)^2 m_{\chi }^4-4 \mu ^2 m_{\chi }^2 \left(\mu ^2 s+s+t\right)+\mu ^4 \left(2 s^2+2 s t+t^2\right)}{\mu^4}$ \\  \hline
  D8 & $\bar \chi \gamma_\mu \gamma^5 \chi\; \bar f \gamma^\mu \gamma^5 f $ & ${1}/{\Lambda_f^2}$ &  $2\frac{(c_N^A)^2}{\Lambda_f^4} \frac{2 \left(\mu ^4+10 \mu ^2+1\right) m_{\chi }^4-4 \left(\mu ^2+1\right) \mu ^2
   m_{\chi }^2 (s+t)+\mu ^4 \left(2 s^2+2 s t+t^2\right)}{\mu ^4}$ \\  \hline
  D9 & $\bar \chi \sigma_{\mu\nu} \chi\; \bar f \sigma^{\mu\nu} f $ & ${1}/{\Lambda_f^2}$ & $8\frac{(c_N^T)^2 }{\Lambda_f^4} \frac{4 \left(\mu ^4+4 \mu ^2+1\right) m_{\chi }^4-2 \left(\mu ^2+1\right) \mu ^2 m_{\chi
   }^2 (4 s+t)+\mu ^4 (2 s+t)^2}{\mu ^4}$  \\  \hline
 D10 & $\bar \chi \sigma_{\mu\nu} \gamma^5\chi\; \bar f \sigma^{\mu\nu} f \;$ & ${i}/{\Lambda_f^2}$ &  $8\frac{(c_N^T)^2 }{\Lambda_f^4} \frac{4 \left(\mu ^2-1\right)^2 m_{\chi }^4-2 \left(\mu ^2+1\right) \mu ^2 m_{\chi }^2 (4 s+t)+\mu ^4 (2 s+t)^2}{\mu^4}$ \\  \hline
\end{tabular}}
\caption{Dimension 6 EFT operators and squared matrix elements for the scattering of Dirac DM from SM fermions \cite{Goodman:2010ku}. The effective couplings for each operator are given as a function of the fermion Yukawa coupling, $y_f$, $\Lambda_f$ is the EFT cutoff scale, and $\mu=m_\chi/m_f$. The fourth column shows the squared matrix elements at high energy as a function of the Mandelstam variables $s$ and $t$. The coefficients $c_N^S$, $c_N^P$, $c_N^V$, $c_N^A$ and $c_N^T$ are the nucleon couplings and are given in appendix~\ref{sec:operators}.  
\label{tab:operatorshe}}
\end{table}

As a first approximation to calculate the capture rate, 
we neglect the temperature of the ion targets. 
In the limit $T_\star\rightarrow0$, the interaction rate $\Omega^-$ can be approximated by (for further details see appendix~C of ref.~\cite{Busoni:2017mhe})
\begin{equation}
\Omega^-(w) = \frac{4\mu_+^2}{\mu w}n_T(r)\int_{w\frac{|\mu_-|}{\mu_+}}^{v_{esc}}dv v  \frac{d\sigma_{T\chi}}{d\cos\theta}.
\end{equation}
In order to calculate the  DM-ion cross section $\sigma_{T\chi}$ in a model independent manner, we assume that the interactions of fermionic DM with quarks 
are described by dimension-6 effective operators, parametrized by a cutoff scale $\Lambda_q$, as listed in Table~\ref{tab:operatorshe}. 
Next, we decompose them in the basis of non-relativistic (NR)  operators $M_N^{\rm NR} = C_i^N {\cal O}_i^{{\rm NR}}$ as outlined in ref.~\citep{DelNobile:2013sia}, and presented here in Table~\ref{tab:operatorsle}, where $N=p,n$ and $C_i$ are the coefficients that accompany the corresponding NR operators. 
Note that since we consider  WDs composed solely of  one of the following elements, $^{4}$He, $^{12}$C or $^{16}$O,  the only operators with non-zero DM-nucleon scattering amplitudes are those with spin-independent interactions, namely D1, D2, D5, D6 and D10.
We then use these coefficients, and the nuclear response functions given in appendix~C of ref.~\cite{Catena:2015uha} (calculated for DM capture in the Sun), to obtain the DM-ion scattering amplitude
\begin{equation}
\MsqT= \frac{m_T^2}{m_N^2}\sum_{i,j} \sum_{N,N'} C_i^N C_j^{N'} F_{i,j}^{(N,N')}.
\end{equation}
Note that the form factors $F_{i,j}^{(N,N')}$ are a function of the transferred momentum, the DM-nucleus relative velocity, $\mu$ and $m_T$. 
The DM-ion cross section then reads
\begin{equation}
\frac{d\sigma_{T\chi}}{d\cos\theta}=\frac{1}{32\pi}\frac{\MsqT}{(m_\chi+m_T)^2}. 
\end{equation}

\begin{table}
\centering
{\renewcommand{\arraystretch}{1.3}
\begin{tabular}{ | c | c | c | c |}
  \hline                        
  Name & Operator & Coupling & $M_N^{\rm NR}$ \\   \hline
  D1 & $\bar\chi  \chi\;\bar N  N $ & $i{c_N^S}/{\Lambda_q^2}$ &  $ 4\frac{ic_N^S}{\Lambda_q^2} m_\chi m_N  {\cal O}_1^{\rm NR}$ \\  \hline  
  D2 & $\bar\chi \gamma^5 \chi\;\bar N N $ & $i{c_N^S}/{\Lambda_q^2}$ &  $-4 \frac{ic_N^S}{\Lambda_q^2} m_N {\cal O}_{11}^{\rm NR}$  \\  \hline  
  D5 & $\bar \chi \gamma_\mu \chi\; \bar N \gamma^\mu N$ & ${c_N^V}/{\Lambda_q^2}$ & $4 \frac{c_N^V}{\Lambda_q^2} m_\chi m_N {\cal O}_1^{\rm NR}$\\  \hline   
  D6 & $\bar\chi \gamma_\mu \gamma^5 \chi\; \bar  N \gamma^\mu N $ & ${c_N^V}/{\Lambda_q^2}$ &   $ 8  \frac{c_N^V}{\Lambda_q^2} (m_\chi m_N  {\cal O}_8^{\rm NR} + m_\chi {\cal O}_9^{\rm NR}) $ \\  \hline   
 D10 & $\bar \chi \sigma_{\mu\nu} \gamma^5\chi\; \bar N \sigma^{\mu\nu} N \;$ & $i{c_N^T}/{\Lambda_q^2}$ &  $ 8 \frac{ic_N^T}{\Lambda_q^2} ( m_\chi {\cal O}_{11}^{\rm NR} - m_N{\cal O}_{10}^{\rm NR} -4 m_\chi m_N{\cal O}_{12}^{\rm NR}) $   \\  \hline
\end{tabular}}
\caption{High energy EFT operators with spin-independent interactions in the non-relativistic limit. The effective couplings are given in terms of the coefficients $c_N^S$, $c_N^V$  and $c_N^T$, and the cutoff scale, $\Lambda_q$. The fourth column shows the matrix elements in the basis of the NR operators given in ref.~\cite{DelNobile:2013sia}. 
\label{tab:operatorsle}}
\end{table}

\begin{figure}
    \centering
    \includegraphics[width=0.49\textwidth]{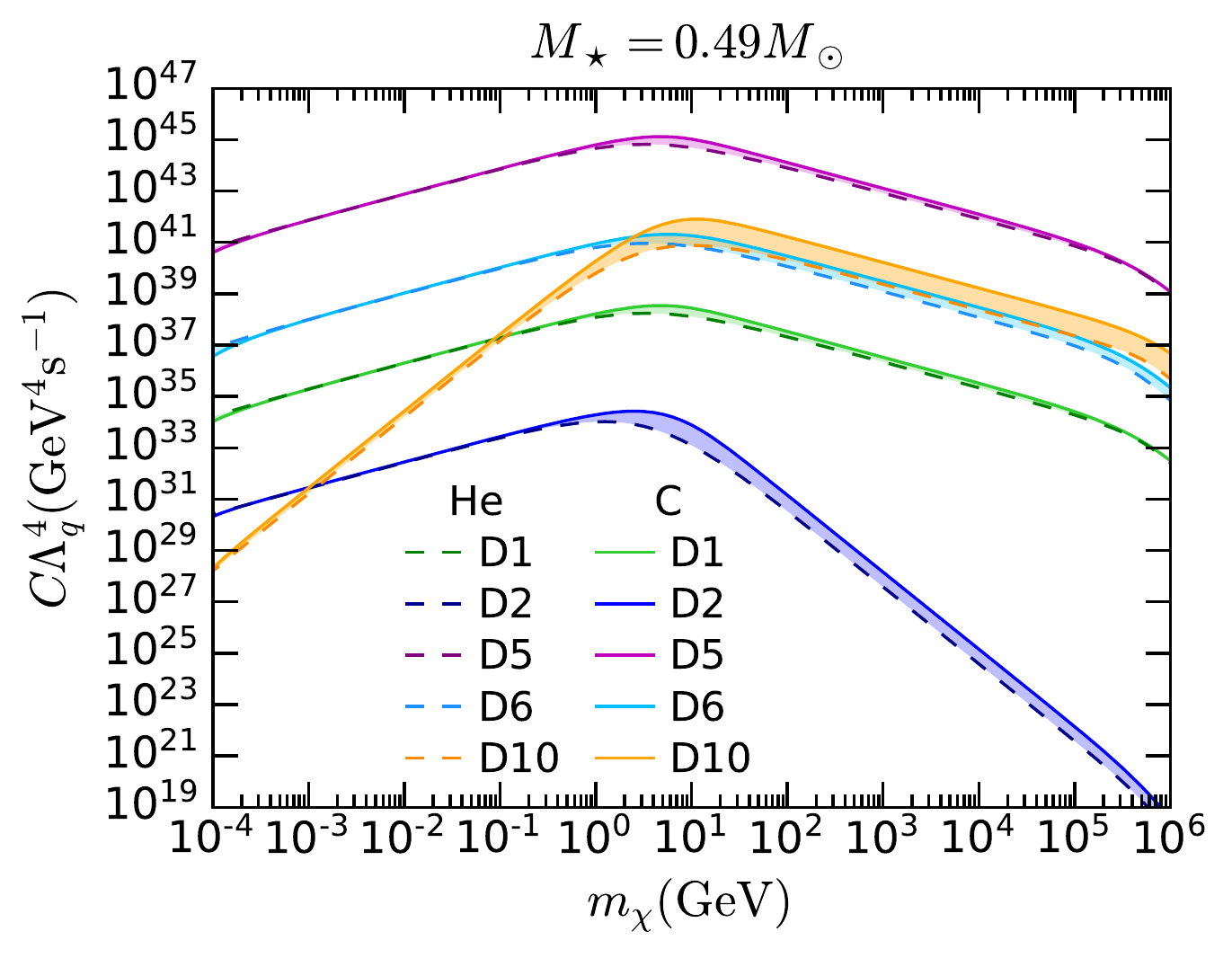}
    \includegraphics[width=0.49\textwidth]{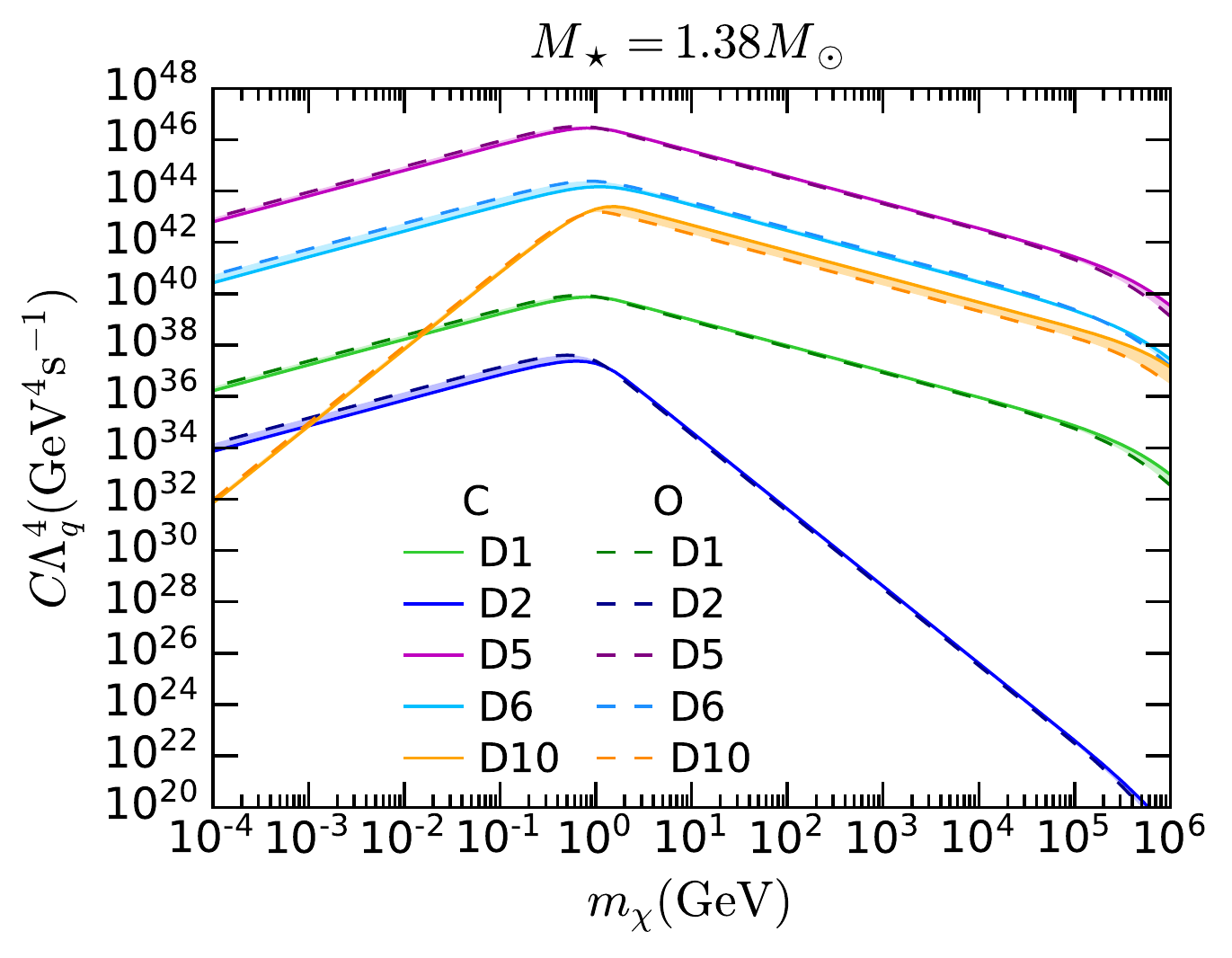}
 \caption{
 Capture rate in the optically thin limit as function of the DM mass for  the lightest (WD$_1$) and heaviest (WD$_4$)  carbon WDs (solid lines) in Table~\ref{tab:WDs}. We also show in the left and right panels the capture rate calculation for WDs made of helium and oxygen (dashed lines), respectively. 
 }
    \label{fig:CoptthinC}
\end{figure}

In Fig.~\ref{fig:CoptthinC} we show the results for $C\Lambda_q^4$ in the  optically thin limit, in the zero temperature approximation, for the lightest (left) and heaviest (right) WDs in Table~\ref{tab:WDs}, as a function of the DM mass. The solid lines correspond to carbon WDs, with the dashed lines representing an alternative composition of helium (oxygen) for the lightest (heaviest) WD. 
The capture rate in the $m_\chi\gtrsim 1\GeV$ mass range is suppressed, for all operators, due to the dependence of the form factor on the momentum transfer. Although heavier DM particles allow for a large range of scattering energies, the larger momentum transfers are suppressed. 
There is another source of suppression, applicable in the  $m_\chi\lesssim 1\GeV$ mass range, which stems from the  conversion  between the DM-nucleon (in the NR limit) to the DM-nucleus cross section. 
The operator D10 features a scaling at low DM that is  substantially different from the other operators, which once again comes from the above mentioned conversion, under which D10 is essentially a momentum suppressed operator in the non-relativistic limit.
At around $m_\chi\sim5\times10^5\GeV$ a second change of slope marks out the region where multiple scattering becomes  relevant to the capture process, but we do not account for these effects here.

The vast majority of WD cores are thought to be made of carbon and oxygen, however their exact composition depends on  the chemical composition of their progenitors, which is reflected in the WD mass. Thus, WDs with masses $\Mstar\lesssim0.5\Msun$  should be comprised of a helium degenerate core, while the main component of heavy WDs, $\Mstar\gtrsim 1.1-1.2\Msun$, could also be oxygen~\cite{Salaris:2018}. To examine how a different WD composition affects the capture rate, in the left panel of Fig.~\ref{fig:CoptthinC} we also plot the capture rate for a $^4$He WD (dashed lines). As expected, less DM will be accreted in WDs made of lighter nuclei. The momentum suppressed operators D10 and D2 feature the largest variation in the capture rate along the whole DM mass range, surpassing the order of magnitude difference at $m_\chi\gtrsim m_{\rm He}$ for D10, followed by D6, D1, and D5 for which the maximum ratio of the two rates (C to He) reaches a factor of $\sim2$. Note that these variations  stem not only from  different target masses but also from distinct nuclear response functions. In the right panel of  Fig.~\ref{fig:CoptthinC}, we compare the capture rate in a WD composed of $^{16}$O (dashed lines)  with the expected rate in a WD of the same mass but instead composed of $^{12}$C. In this case, the higher ratio occurs in the sub-GeV regime, where it reaches a factor of $\sim 2$ for D2 and D6, $\sim 1.4$ for D1 and D5, and $\sim1.3$ for D10. With the sole exception of D10 for $m_\chi\gtrsim1\GeV$, less DM is captured in a carbon WD, 
therefore, as a conservative approximation we can assume that this WD ($\Mstar=1.38\Msun$) and in general WDs with $\Mstar\gtrsim 0.5\Msun$ are entirely made of carbon.  For WDs with masses below this threshold, the conservative assumption would be a He degenerate core.

\subsection{Star Opacity}
\label{sec:ionopacity}

When the scattering cross section is large enough that the capture rate approaches the geometric limit, the so-called threshold cross section, the flux of DM particles is considerably attenuated when traversing the star.  We then enter a regime where the optically thin approximation no longer holds and hence the star opacity should be considered in the capture rate computation. 
As in refs.~\cite{Busoni:2017mhe,Bell:2020jou}, we quantify the WD opacity with the optical factor $\eta(r)$ defined in terms of the optical depth $\optdepth$, i.e., the integral of the interaction rate along the path, $\gamma$, followed by the DM particle until it is finally captured~\cite{Bell:2020jou}
\begin{eqnarray}
\eta(\optdepth)&=&e^{-\optdepth}, 
\label{eq:optfactor}\\
    \tau_\chi(r)&=&\int_\gamma\Omega^-(r)\frac{d\tau}{dr}dr,
\end{eqnarray}
where  $\tau$ is the proper time. The $\eta$ factor removes DM particles from the incoming flux once they are captured,  avoiding double-counting of interactions. 

To evaluate the optical depth $\optdepth$, we follow the improved approach that was developed in ref.~\cite{Bell:2020jou} for neutron stars. This approach correctly incorporates the shapes of the orbits DM particles follow inside the star. There are two possible trajectories a DM particle could follow from the surface of the star to a point $x$ in a shell of radius $r$ within the WD. Unlike the shortest path, the longest goes across the perihelion before reaching $x$.  These equally probable trajectories lead to two optical depths $\tau^-_\chi$ and $\tau^+_\chi$, respectively, which are defined as~\cite{Bell:2020jou}  
\begin{eqnarray}
    \tau^-_\chi(r,J)
     &=&\int_r^{R_\star}\frac{dx}{\sqrt{1-\frac{J^2}{J_{max}(x)^2}}}\frac{\Omega^-(w(x))}{\sqrt{B(x)(1-B(x))}},\label{eq:tauminusdef}\\
     \tau^+_\chi(r,J)
    &=&\tau^-_\chi(r_{min},J)+2\int_{r_{min}}^r\frac{dx}{\sqrt{1-\frac{J^2}{J_{max}(x)^2}}}\frac{\Omega^-(x)}{\sqrt{B(x)(1-B(x))}},\label{eq:tauplusdef}
\end{eqnarray}
 where $J$ is the DM angular momentum, $J_{max}(x)$ the maximum value it can obtain at a distance $x$ from the centre of the star, $r_{min}$ is the position of the perihelion and $B$ is the factor that appears in the time component of the Schwarzschild metric. In WDs, this metric factor,  which encodes general relativity (GR) corrections, is very close to 1. It is, therefore, convenient to expand $B$ in terms of the escape velocity as
\begin{equation}
    B(r) \sim 1 - v_{esc}^2(r).
\end{equation}
Note that given that $v_{esc}(r)  \gg {\rm Max}[\vstar,v_d]$ (see Fig.~\ref{fig:WDradprofs}, bottom panel.), we can safely neglect the DM initial velocity $u_\chi$.

The angular momentum $J$ can be recast in terms of its maximum value, 
  \begin{equation}
  J=yJ_{max}(r) = y m_\chi r w, \qquad 0\leqslant y \leqslant 1.
  \end{equation}
Then, averaging over the possible trajectories and the DM angular momentum, the optical factor is given by 
\begin{equation}
    \eta(r)=\frac{1}{2}\int_0^1\frac{y dy}{\sqrt{1-y^2}}\left(e^{-\tau_{-}(r,y)}+e^{-\tau_{+}(r,y)}\right).\label{eq:eta}
\end{equation}
Finally, to account for the WD opacity, we introduce the optical factor $\eta(r)$, Eq.~\ref{eq:eta}, into  Eq.~\ref{eq:ioncapdef} to obtain
\begin{equation}
     C_{opt} = \frac{\rho_\chi}{m_\chi}\int_0^{R_\star} 4\pi r^2\eta(r)     
     \int_0^\infty du_\chi \frac{w(r)}{u_\chi} \fMB(u_\chi) \Omega^-(w).\label{eq:capoptics}     
\end{equation}

\begin{figure}[t]
    \centering
  \includegraphics[width=0.49\textwidth]{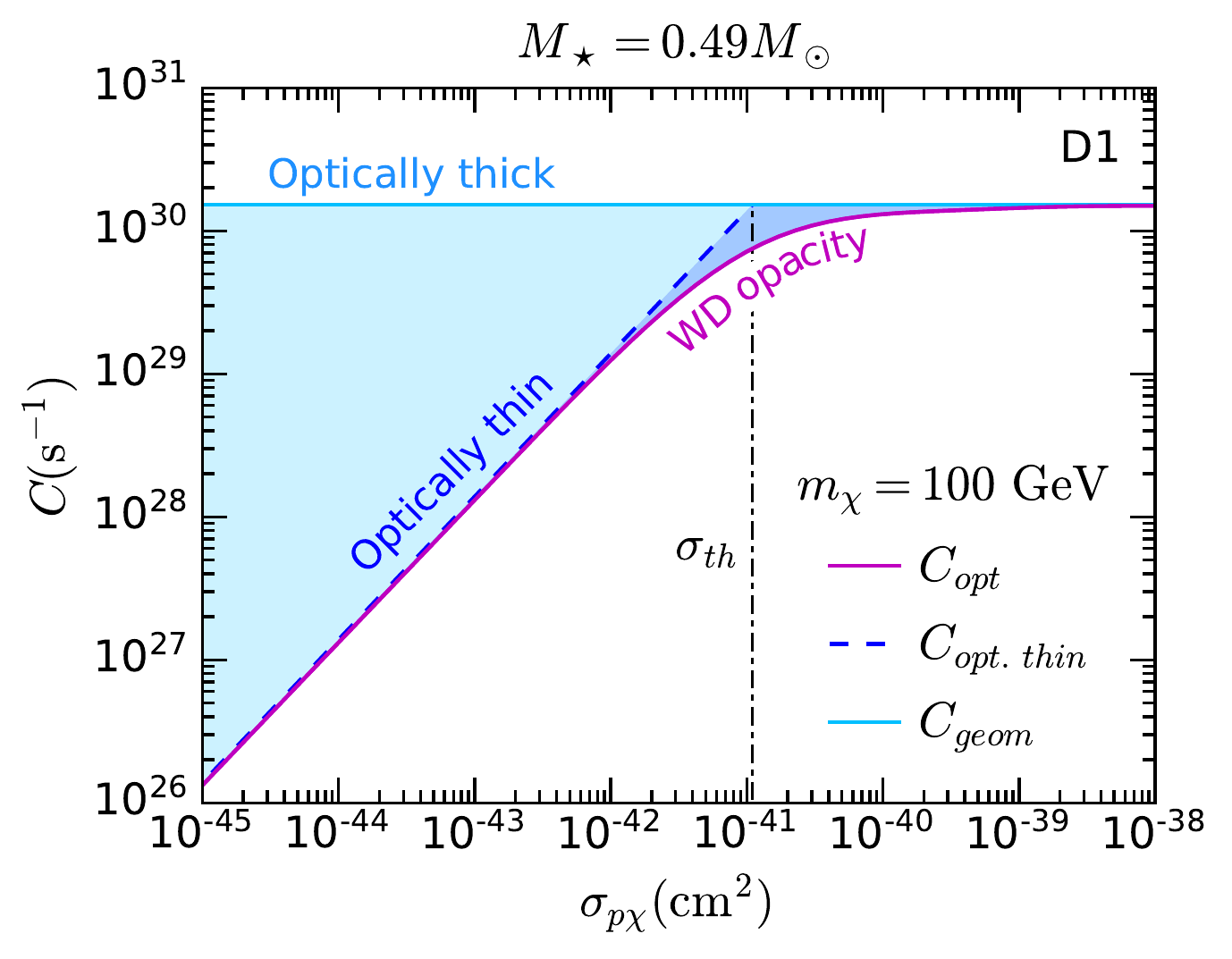}
  \includegraphics[width=0.49\textwidth]{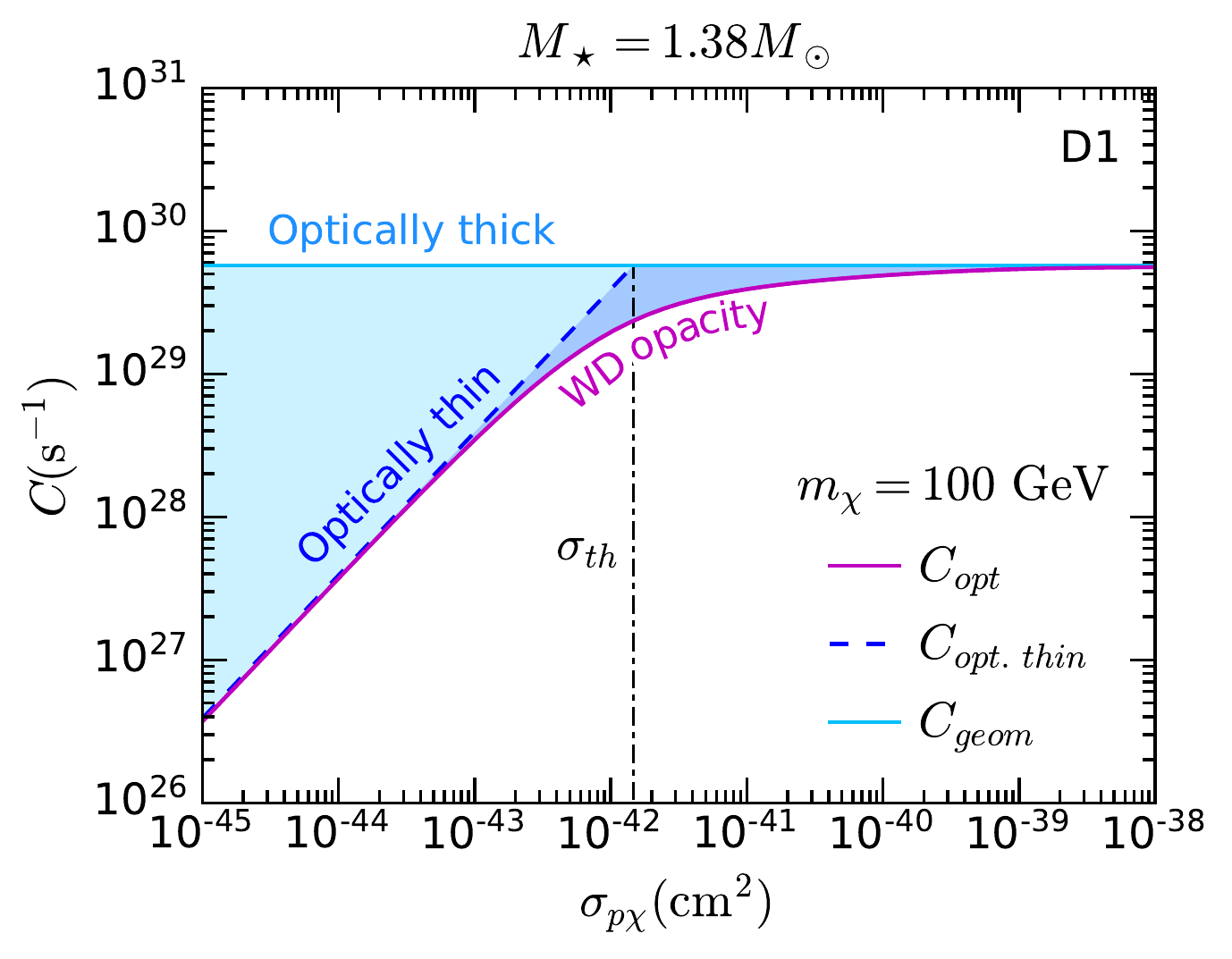} 
    \caption{Capture rate for the carbon WDs in Fig.~\ref{fig:CoptthinC}, as a function of the DM-proton cross section for the scalar-scalar operator and $m_\chi=100\GeV$. The magenta line denotes the calculations that accounts for the star opacity. The geometric limit (optically thick regime) is shown in light blue and the dashed blue line indicates the optically thin limit $\eta(r)=1$.  The dot-dashed black line represents the threshold cross section $\sigmath$.}
    \label{fig:opacity}
\end{figure}

In Fig.~\ref{fig:opacity} we show the capture rate calculated with Eq.~\ref{eq:capoptics} (solid magenta) as a function of the DM-proton cross section,  using the same WDs as in Fig.~\ref{fig:CoptthinC}, assuming $m_\chi=100\GeV$ and scalar-scalar interactions (D1). We can observe that the calculation that includes the star opacity realises the transition from the optically thin (dashed blue) to the optically thick (geometric) limit depicted in light blue.  Note that the maximum  DM-proton cross section for which the optically thin limit is a good approximation varies from $\sigma_{p\chi}\sim 10^{-41}\cm^2$  for a carbon WD in M4 with $\Mstar=0.49\Msun$ to $\sigma_{p\chi}\sim 10^{-42}\cm^2$  for  $\Mstar=1.38\Msun$. For larger values of $\sigma_{p\chi}$ the optical factor $\eta(r)$ increasingly suppresses the capture rate so that it is no longer proportional to $\sigma_{p\chi}$ and saturates to the geometric limit. As in ref.~\cite{Bell:2020jou}, we can define the threshold cross section $\sigmath$ as the intersection of the optically thin and geometric limits (see dot-dashed black line), determining the maximum $\sigma_{p\chi}$  for which the optical thin limit can be used.

\subsection{Finite temperature and Evaporation}
\label{sec:evapions}

\begin{figure}
    \centering
  \includegraphics[width=0.49\textwidth]{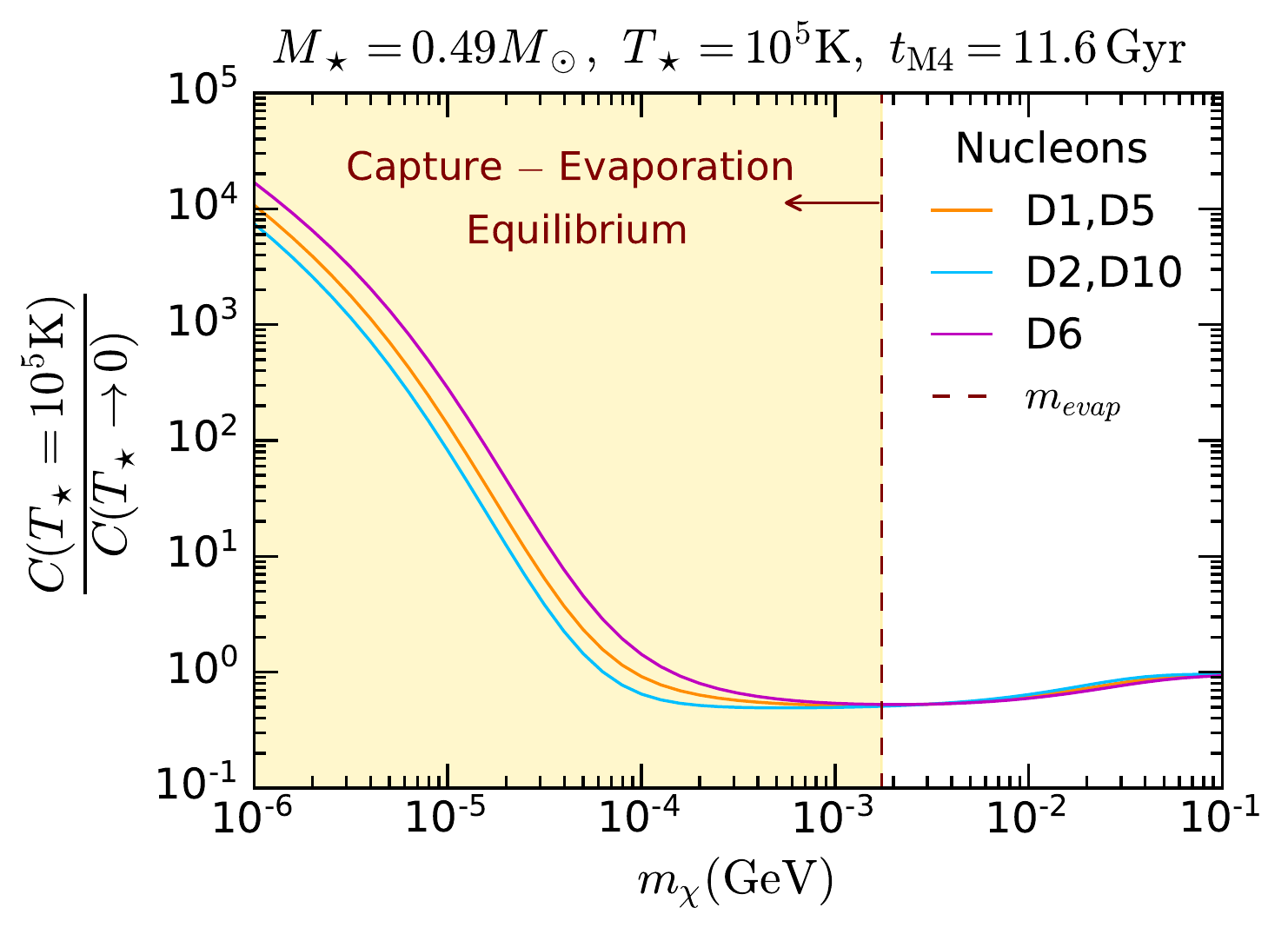}
  \includegraphics[width=0.49\textwidth]{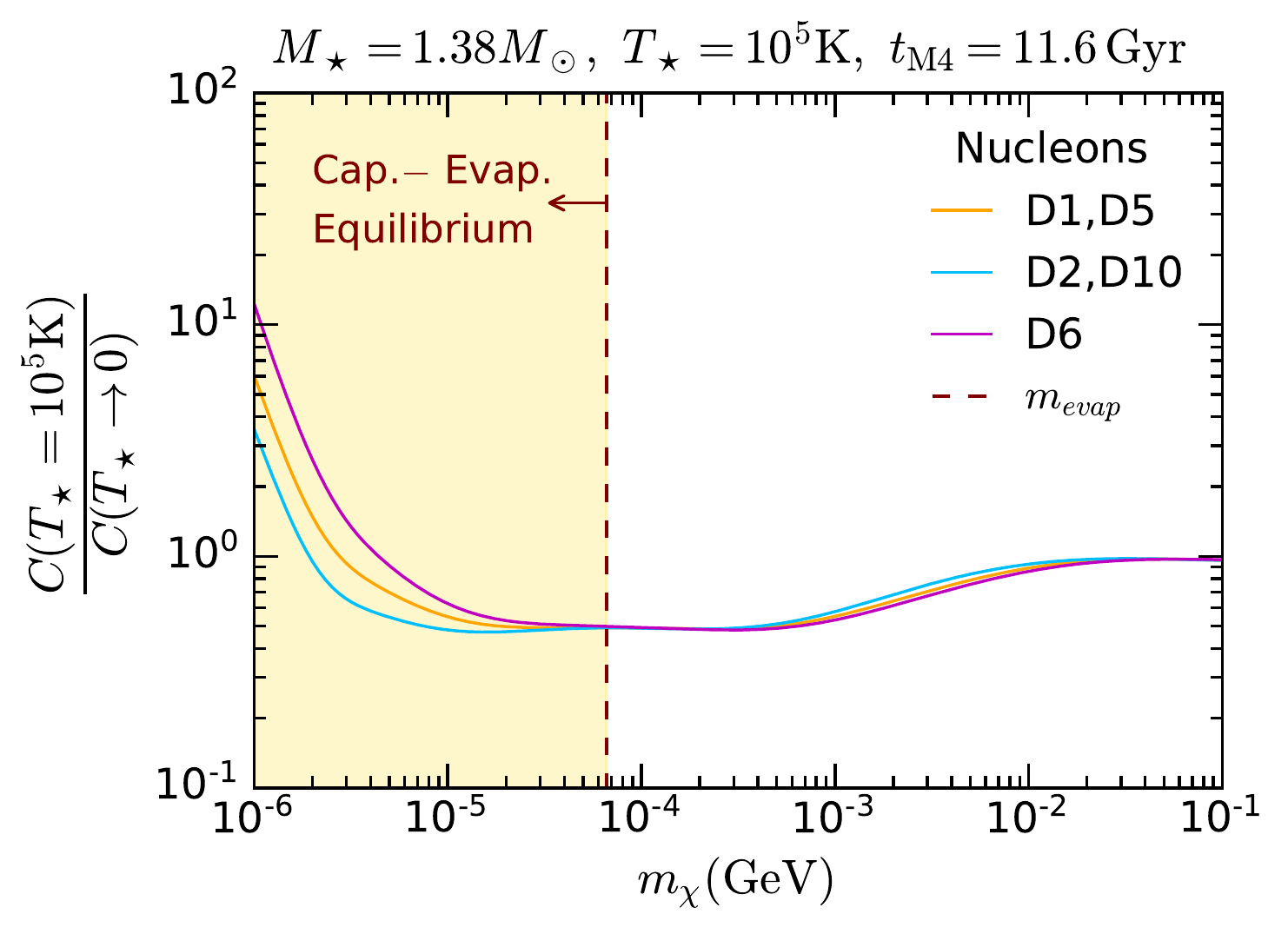} 
    \caption{Finite temperature effects on the capture rate in the case of scattering on carbon targets, for two WDs in the globular cluster M4, namely WD$_1$ (left) and WD$_4$ (right) of Table~\ref{tab:WDs}. The DM mass range where capture and evaporation are expected to be in equilibrium is shaded in yellow. The dashed brown line corresponds to the evaporation mass. }
    \label{fig:capTfin}
\end{figure}

For WDs in globular clusters, where the DM velocity dispersion is expected to be low ($v_d=8\km\s^{-1}$ for M4) in comparison with DM falling into stars located in the solar neighbourhood ($v_d=270\km\s^{-1}$), finite temperature effects are expected to be important as long as the thermal speed of the target $v_T=\sqrt{\Tstar/m_T}$  is greater than the velocity dispersion $v_d$ (see Eq.~\ref{eq:vdfinT}). For instance, at $\Tstar=10^6\K$, for which $v_T\simeq26\km\s^{-1}$, this high temperature induces a flattening of the DM-target relative speed distribution, Eq.~\ref{eq:veldistfiniteT}. 
Consequently, the $\Tstar\rightarrow0$ limit is no longer a good approximation for this core temperature, and Eqs.~\ref{eq:veldistfiniteT} and \ref{eq:ionRdef} should be use to calculate the capture rate.

On the other hand, in order to determine the  core temperatures of observed WDs, we need to know the atmospheric composition, especially the hydrogen content.  This is done using spectroscopic observations, which are not available for WDs in M4, and model atmospheres. For the faintest (hence the oldest) and heaviest WDs in M4, a temperature of $\Tstar=10^5\K$ is consistent with the age estimated for the globular cluster, using the evolutionary sequences given in ref.~\cite{Bedard:2020}~\footnote{These evolutionary sequences are publicly available at  \url{http://www.astro.umontreal.ca/~bergeron/CoolingModels/}.} as a reference to estimate $\Tstar$ from observational data.
In Fig.~\ref{fig:capTfin} we plot the ratio of the capture rate with $\Tstar=10^5\K$ to the zero temperature approximation, for two WDs (WD$_1$ and WD$_4$ of Table~\ref{tab:WDs}). For the lightest (heaviest) WD, the ratio starts to deviate from 1 at $m_\chi\sim 100\MeV \, (10 \MeV)$ for all operators and becomes increasingly higher at around $m_\chi\sim 100\keV \, (5\keV)$. The velocity and momentum suppressed operator D6 exhibits the largest correction for very low DM masses, followed by D1, D5, whose cross sections depend on the transferred momentum only through the form factor and D2, D10 which are momentum suppressed.
As expected, temperature corrections are more important in the low mass WD, WD$_1$, where the escape velocity is lower than that for the heavier WD$_4$ by a factor $\gtrsim 5$; see Fig.~\ref{fig:WDradprofs}.

Accreted DM accumulated in the WD core can also up-scatter and potentially evaporate from the star, i.e., drain energy from the target in a collision and acquire a final velocity greater than the local escape velocity. This process affects the very light DM regime.  The formalism employed to calculate the evaporation rate is very similar to that of the capture rate, the key difference being that due to the nature of the evaporation process, the WD temperature cannot be neglected.

To calculate the evaporation rate, we first determine the DM speed distribution within the star. 
As discussed in section~\ref{sec:wd}, the  WD interior is expected to be isothermal with temperature $\Tstar$, except for the outermost layers which are in local thermal equilibrium. Hence, we can neglect the WD envelope, assume that the radius of the sphere where the accreted DM lies is $r_\chi\ll\Rstar$, 
and describe the DM distribution in the core simply with an isothermal profile characterised by a temperature $\Tstar$. 
The DM number density in the WD interior then is given by 
\begin{equation}
n_\chi(r)=n_{\rm 0} e^{-m_{\chi}\phi(r)/\Tstar}\label{eq:niso},
\end{equation} 
where $\phi(r)$ is the gravitational potential and $n_{0}$ is a normalisation constant, defined such that the integral of the DM velocity distribution is normalised to 1.
The DM velocity distribution  within the WD is described by a MB distribution 
\begin{equation}
f_{evap}(v)=n_\chi(r) e^{-\frac{m_{\chi}v^2}{2\Tstar}}\label{eq:dmsundistribsimple}. 
\end{equation}

Using Eq.~(\ref{eq:dmsundistribsimple}),  the evaporation rate reads~\citep{Busoni:2017mhe}
\be
E = \int_0^{\Rstar} dr 4\pi r^2 \eta(r) \int_0^{v_{esc}(r)} dw f_{evap}(w)  \Omega^+(w),
\ee
where the interaction rate for up-scattering $\Omega^{+}(w)$ is defined as
\begin{eqnarray}
\Omega^{+}(w) &=& \int_{v_e}^\infty R^{+}(w\rightarrow v)  d v, \\
R^{+}(w\rightarrow v) &=& \int_0^\infty ds' \int_0^\infty d t' \frac{32 \mu_{+}^4}{\sqrt{\pi}}\kappa^3 n_T(r) \frac{d\sigma_{T\chi}}{d\cos\theta} \frac{v t'}{w} e^{-\kappa^2v_T^2} \Theta(t'+s'-v)\Theta(w-|t'-s'|).
\end{eqnarray}
It is worth emphasising that, in the computation of the evaporation rate, the $\Tstar\rightarrow0$ approximation does not hold since the DM gains energy from the targets.

Evaporation will alter the rate at which the DM is accumulated within the star, such that 
\begin{equation}
    \frac{dN_\chi}{dt} = C -EN_\chi,\label{eq:evap_boltz}
\end{equation}
where $N_\chi$ is the number of DM particles within the star. Then, assuming that the capture and evaporation rates remain constant over time, the  number of DM particles accumulated throughout the lifetime of the star is 
\begin{equation}
    N_\chi(t_\star) = C t_\star\left(\frac{1 - e^{-E t_\star}}{E t_\star}\right), \label{eq:DM_num}
\end{equation}
where $\tstar$ is the age of the WD. 
From this equation, one can see that the effect of evaporation is negligible unless $E t_\star \gtrsim\mathcal{O}(1)$. Therefore, we can estimate the evaporation mass $m_{evap}$ as the DM mass for which $E(m_\chi) t_\star\sim 1$.\footnote{We are using a simplified approach to estimate the evaporation mass. To be exact, we should include the annihilation term in Eq.~\ref{eq:evap_boltz} and calculate $m_{evap}$ at $E(m_\chi)\tau_{eq} \sim 1$, where $\tau_{eq}$ is the timescale for capture-annihilation equilibrium defined in Eq.~\ref{eq:taueq}. Our assumption relies on the fact that the evaporation mass is determined by the exponential factor arising from the evaporation rate. The maximum error in the evaporation mass introduced by this approximation is $\sim50\%$, when the annihilation rate is high.}  Aside from the age of the star, this mass threshold depends on the DM-target cross section and heavily on the WD core temperature. While an order of magnitude variation in the cross section changes  $m_{evap}$ by a factor of $\sim 1.5-1.9$, a similar change in the star temperature translates into about an order of magnitude variation in $m_{evap}$. Also note that the evaporation mass is approximately independent of the operator considered.

In Fig.~\ref{fig:capTfin}, we show our estimates of the evaporation mass (dashed brown lines) for two WDs located in the globular cluster M4. We adopt a core temperature of $\Tstar=10^5\K$, and assume that the WDs are the same age as the globular cluster, i.e. $\tstar=t_{\rm M4}$.\footnote{Note however that during the first Gyrs of life, WDs are much hotter and that the rate at which these compact stars cool down depends heavily on the atmosphere composition. Consequently, a more accurate estimation of $m_{evap}$ requires a simulation of the WD evolution, which is beyond the scope of this work.} 
This is a conservative assumption since the evaporation mass is lower in younger WDs. We find $m_{evap}\simeq1.7\MeV$ (left) and $m_{evap}\simeq65\keV$ (right) for carbon WDs with $\Mstar=0.49\Msun$ and $\Mstar=1.38\Msun$, respectively, illustrating the hierarchy in $m_{evap}$ for different WD configurations. Note also that finite  temperature effects on the capture rate come into play before the evaporation mass is reached, suppressing $C$ up to a factor of $\sim50\%$. Below the evaporation mass, the capture and evaporation processes are in equilibrium and hence we expect no net energy exchange due to DM.

\section{Capture by Scattering on Electrons}
\label{sec:capelectrons}
Unlike ions, electrons in WDs are highly degenerate and, in the case of massive WDs, are relativistic. Therefore, in order to calculate the capture rate, we require a formalism that properly incorporates Pauli blocking and relativistic kinematics, together with the internal structure of the WD. 
We thus adopt the formalism developed in refs.~\cite{Bell:2020jou, Bell:2020lmm} for the case of DM capture in neutron stars, which includes the effects mentioned above, uses relativistic kinematics, and also considers GR corrections.  In the case of WDs, the latter are small.

\subsection{Single scattering}
The expression for the capture rate derived in refs.~\citep{Bell:2020jou, Bell:2020lmm} assumed a MB speed distribution and neglected the DM velocity far away from the star, as it is small compared to the escape velocity. While the same approximation holds for WDs, one might be interested in departures from the standard MB speed  distribution. 
The generalised expression for the capture rate without integrating over the DM velocity $u_\chi$ is 
\begin{equation}
C = 4\pi \frac{\rho_\chi}{m_\chi} \int_0^\infty \frac{\fMB(u_\chi)du_\chi}{u_\chi}
\times\int_0^{\Rstar} \eta(r) r^2 \frac{\sqrt{1-B(r)}}{B(r)} \Omega^{-}(r)  \, dr,     \label{eq:capturelec}
\end{equation}
where the opacity factor $\eta(r)$  can be taken to be $1$ when $\sigma_{e\chi}\ll\sigma_{e\chi}^{th}\sim 10^{-39}\cm^2$. 
The rate of DM interactions with electron targets is given by~\cite{Bell:2020jou, Bell:2020lmm} 
\begin{eqnarray}
\Omega^{-}(r) &=& \frac{\zeta(r)}{32\pi^3}\int dt dE_e ds  
\frac{|\overline{M}_{e\chi}|^2}{2s\beta(s)-\gamma^2(s)}  \frac{E_e}{m_\chi}\sqrt{\frac{B(r)}{1-B(r)}} \frac{s}{\gamma(s)}\fFD(E_e,r)(1-\fFD(E_e^{'},r))\nn\\
&\times&\Theta\left(E_e\sqrt{\frac{1-B(r)}{E_e^2-m_e^2}}-\frac{s_{max}+s_{min}-2s}{s_{max}-s_{min}}\right)\Theta\left(E_e^{'}-E_e\right),\label{eq:omegampauliur}\\
    \beta(s) &=& s-\left(m_e^2+m_\chi^2\right),\\
    \gamma(s) &=& \sqrt{\beta^2(s)-4m_e^2m_\chi^2},
\end{eqnarray}
where $s$ and $t$ are the  Mandelstam variables and $E_e$ and $E_e^{'}$ are the target electron initial and final energies, respectively. The correction factor $\zeta(r)=\frac{n_{e}(r)}{n_{free}(r)}$ accounts for the fact we are using realistic profiles for the electron number density $n_e(r)$ and the chemical potential $\muFe(r)$, while the interaction rate is defined in the free Fermi gas approximation~\cite{Garani:2018kkd,Bell:2020jou,Bell:2020obw}. 
The definition of $n_{free}(r)$ can be found in ref.~\cite{Bell:2020jou}. 
The integration intervals in Eq.~\ref{eq:omegampauliur} are
\begin{eqnarray}
t_{max} &=& 0, \qquad \qquad
t_{min} = -\frac{\gamma^2(s)}{s},\\
s_{min} &=& m_e^2+m_\chi^2 + 2\frac{E_e m_\chi}{\sqrt{B(r)}}-2\sqrt{\frac{1-B(r)}{B(r)}}m_\chi\sqrt{E_e^2-m_e^2},\label{eq:smin}\\
s_{max} &=& m_e^2+m_\chi^2 + 2\frac{E_e m_\chi}{\sqrt{B(r)}}+2\sqrt{\frac{1-B(r)}{B(r)}}m_\chi\sqrt{E_e^2-m_e^2}\label{eq:smax},\\
E_{e,min} &=& m_e,
\end{eqnarray}
while $E_{e,max}$ should be set to $E_{e,max}=m_e+\muFe$ in the $T_\star\rightarrow0$ limit, or left free otherwise.

We have introduced two additional $\Theta$ functions in Eq.~\ref{eq:omegampauliur} when compared to refs.~\cite{Bell:2020jou, Bell:2020lmm}. 
The first $\Theta$ function ensures that we count only scatterings that are kinematically allowed, i.e., that the collision is head-on. 
We explain the details of the derivation of this phase space constraint in appendix~\ref{sec:phasespace}. 
The second Heaviside function enforces that DM actually loses energy, which is required for finite temperature calculations. 
In the zero temperature  limit, on the other hand, the FD distributions can themselves be approximated by $\Theta$ functions. Therefore, the initial states occupy all the lower energy levels, and scattering can proceed only if the target acquires enough energy to be ejected from the Fermi sea. 
Specifically, the  inequalities enforced are
\begin{eqnarray}
E_e \le m_e +\muFe, \qquad \qquad 
E_e^{'} > m_e +\muFe.
\end{eqnarray}

Finally, note that we do not integrate the interaction rate over the initial DM  speed at infinity in Eq.~\ref{eq:capturelec}, i.e., we do not include the initial DM energy in the interaction rate. This is due to the fact that, when computing the capture rate for WDs in M4,  we find there is no significant impact when correcting for the DM velocity at infinity, even for the lightest WD we consider. This can be understood by noting that the halo velocities are of order $u_\chi^2\sim 10^{-6}$, while the escape velocity is $v_{esc}^2 = 1 - B(r) \sim 10^{-3}$, and so the corrections are expected to be only of order $u_\chi^2/v_e^2  \sim 10^{-3}$.

\subsection{Multiple scattering}
\label{sec:mselectrons}

For DM masses larger than a certain threshold denoted $m_e^*$, and cross sections $\sigma_{e\chi}\ll \sigma_{e\chi}^{th}$, the capture probability for single scattering is no longer $\sim 1$. In this regime, multiple collisions are required in order for the DM particles to lose sufficient energy to be captured. 
In the $T_\star\rightarrow0$ limit, one can use the multiple scattering approach of ref.~\citep{Bell:2020jou}, which involves  inserting the capture probability $c_1(r)$ in Eq.~\ref{eq:omegampauliur} instead of the $\Theta\left( E_e^{'} - E_e\right)$ term, with
\begin{equation}
c_1(r)=\frac{1}{n^*(r)} = 1-e^{-\frac{m^*_e(r)}{m_\chi}}\sim \frac{m^*_e(r)}{m_\chi}, 
\end{equation}
where $n^*(r)$ represents the number of interactions required to capture the incoming DM particle. 

Since the DM energy loss for scattering on electrons 
is much larger than the WD core temperature, as is the case for DM scattering in neutron stars, we can follow refs.~\citep{Bell:2020jou,Bell:2020lmm} to calculate $m^*_e$.  We assume a MB velocity distribution in M4, with $v_\star=20\km\s^{-1}$, $v_d=8\km\s^{-1}$. As an example, if we take a WD of mass $\Mstar=1.38\Msun$, a constant matrix element, $B=0.995$ and $\muFe=8\MeV$, we find 
\begin{equation}
m^*_e \simeq 10^5 \GeV. 
\end{equation}

\begin{figure}
    \centering
    \includegraphics[width=0.49\textwidth]{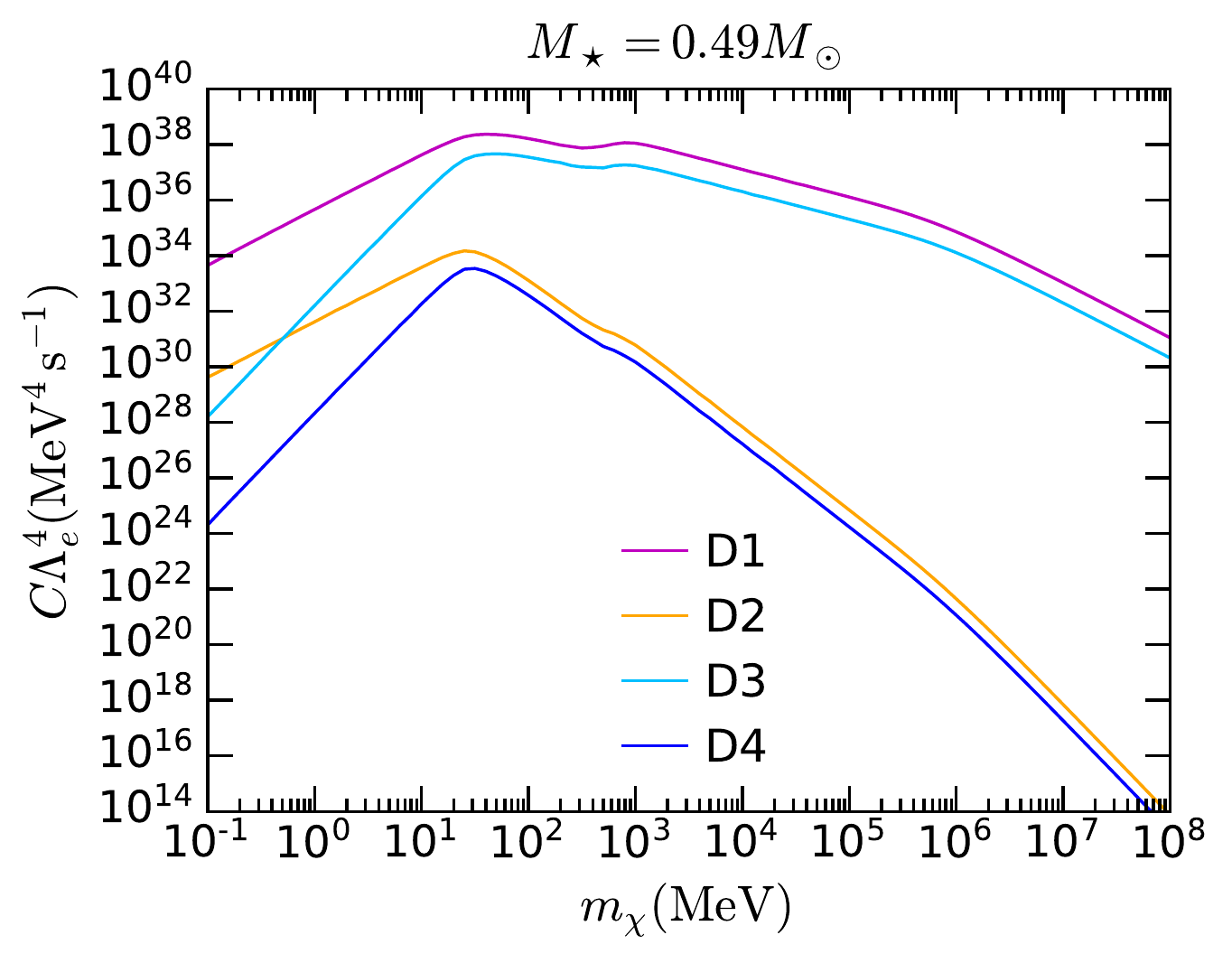}
    \includegraphics[width=0.49\textwidth]{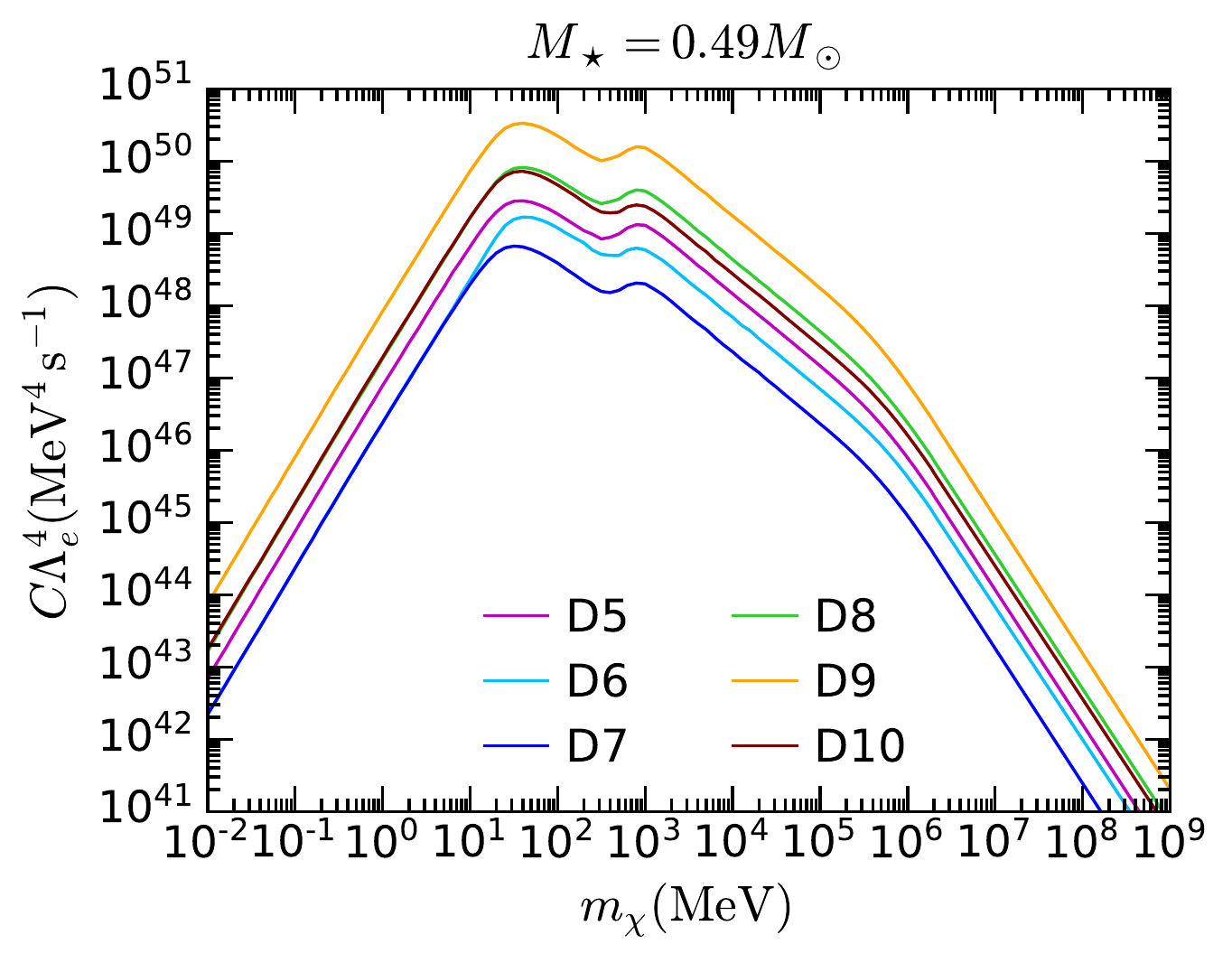} \\    
    \includegraphics[width=0.49\textwidth]{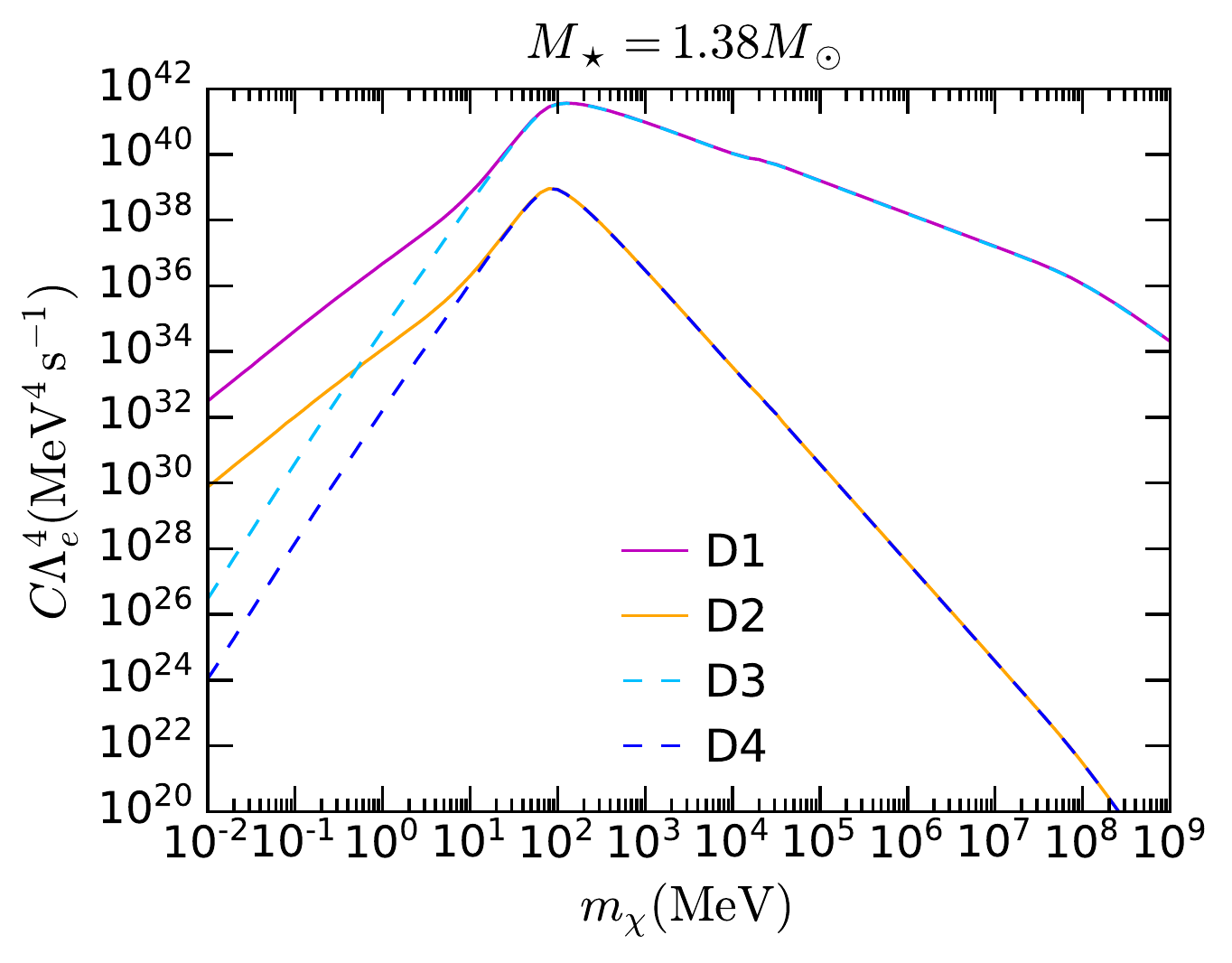}
    \includegraphics[width=0.49\textwidth]{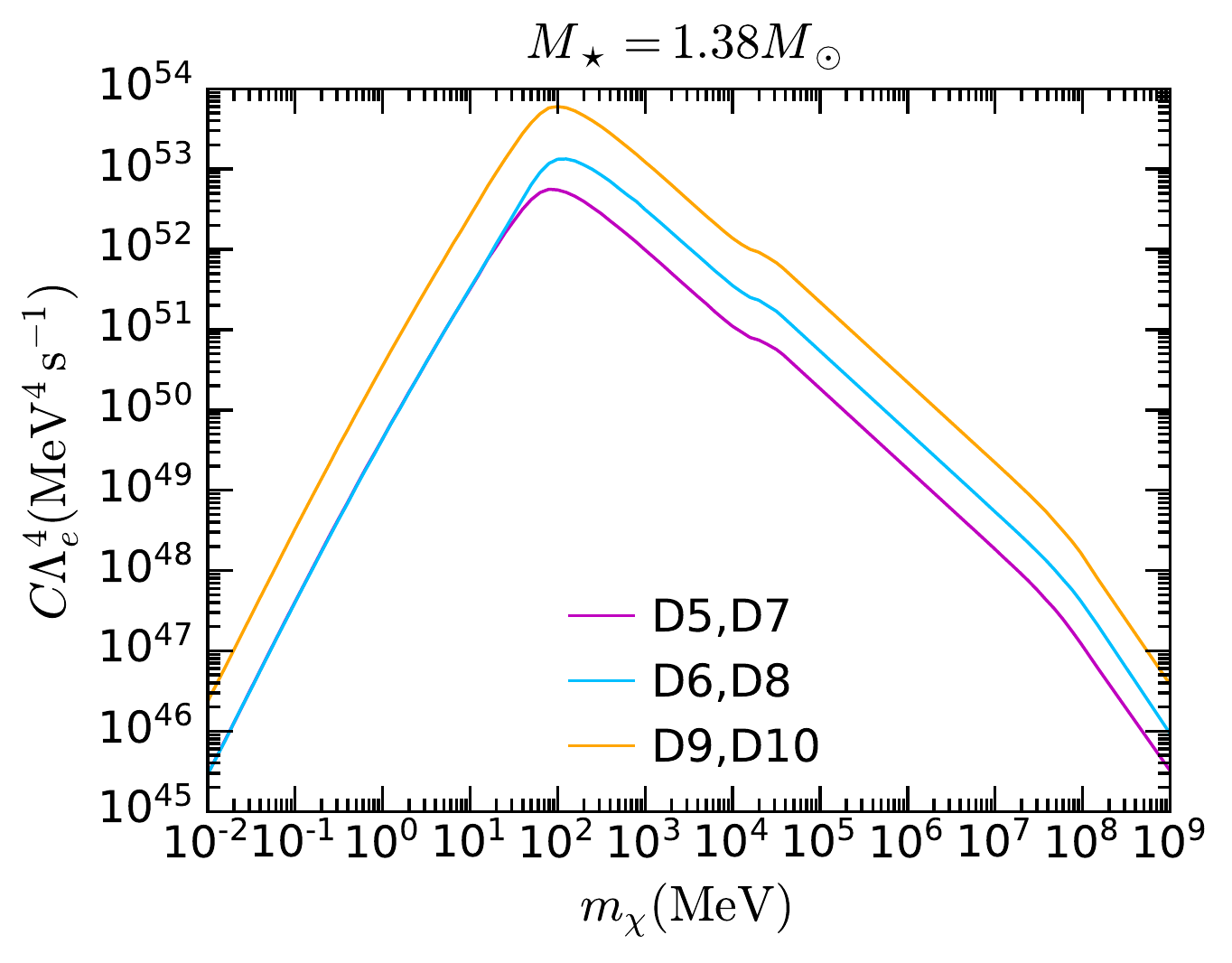} \caption{
 Capture rate for scattering on electrons, in the optically thin limit, as function of the DM mass for  the lightest (WD$_1$, top panels) and heaviest (WD$_4$, bottom panels)  
 carbon WDs in Table \ref{tab:WDs}.}
    \label{fig:Coptthinelec}
\end{figure}

\medskip

We are now ready to calculate the capture rate  for the operators in Table~\ref{tab:operatorshe}, which holds for DM-electron interactions with $\Lambda_f=\Lambda_e$, $\mu=m_\chi/m_e$ and the coefficients $c_N^I$, $I=S,P,V,A,T$ set to 1. Note that the cutoff scale for DM interactions with quarks ($\Lambda_q$) and electrons ($\Lambda_e$) are not necessarily equal. 
In Fig.~\ref{fig:Coptthinelec}, we present our results for $C\Lambda_e^4$ in the zero temperature and optically thin limits for the same WDs we considered in the previous section, namely carbon WDs with $\Mstar=0.49\Msun$ (top panels) and  $\Mstar=1.38\Msun$ (bottom panels). In both WDs, Pauli blocking strongly suppresses the capture rate in the light mass regime,  $m_\chi\lesssim100\MeV$. Above this mass range, Pauli suppression persists but remains minimal.   
The change of slope in the Pauli suppressed region for operators D1 and D2 is due to the fact that their matrix elements contain a factor $(t-4m_e^2)$, which introduces an additional suppression due to the smallness of the electron mass in the $m_e\lesssim m_\chi\lesssim100\MeV$ interval~\cite{Bell:2020lmm}. 
Then, a transition between the Pauli blocked and the non-suppressed capture rate is observed for all the operators, which is immediately noticeable in the case of the light WD, where we observe a valley  in the  $100\MeV\lesssim m_\chi\lesssim1\GeV$ mass range. In the heavy WD, this transition region extends up to $\sim10\GeV$ and is more evident for operators D5-D10. 
The region at which multiple scattering becomes relevant also depends on the star configuration.  It occurs at $m_\chi\gtrsim1\TeV$ for the light WD, 
and at around  $m_\chi\gtrsim10^5\GeV$ for the heavy WD, observed as a change of slope in the capture rate at around those masses. 

It is worth remarking that, in contrast to the light WD, the heavy WD features an electron chemical potential more than one order magnitude higher, and electrons which are relativistic. As a result, the capture rate curves for WD$_4$ exhibit similar features to those observed in neutron stars where electrons are degenerate and ultra-relativistic~\cite{Bell:2020lmm}. In addition, since the electrons in the heavy WD are relativistic, the scattering amplitudes are dominated by terms of the form $t^ns^m$  in the large DM mass regime, while terms proportional to $m_e^2$ are suppressed (see Table~\ref{tab:operatorshe}). This results in very similar capture rates for operators D1 and D3, D2 and D4, D5 and D7, D6 and D8, D9 and D10.  
Finally, contrary to the case of scattering on ions, the capture rate due to scattering on electrons would scarcely be affected by a different chemical composition, such as He or O.

\subsection{Finite temperature and Evaporation}
\label{evapelectrons}
\begin{figure}
    \centering
  \includegraphics[width=0.495\textwidth]{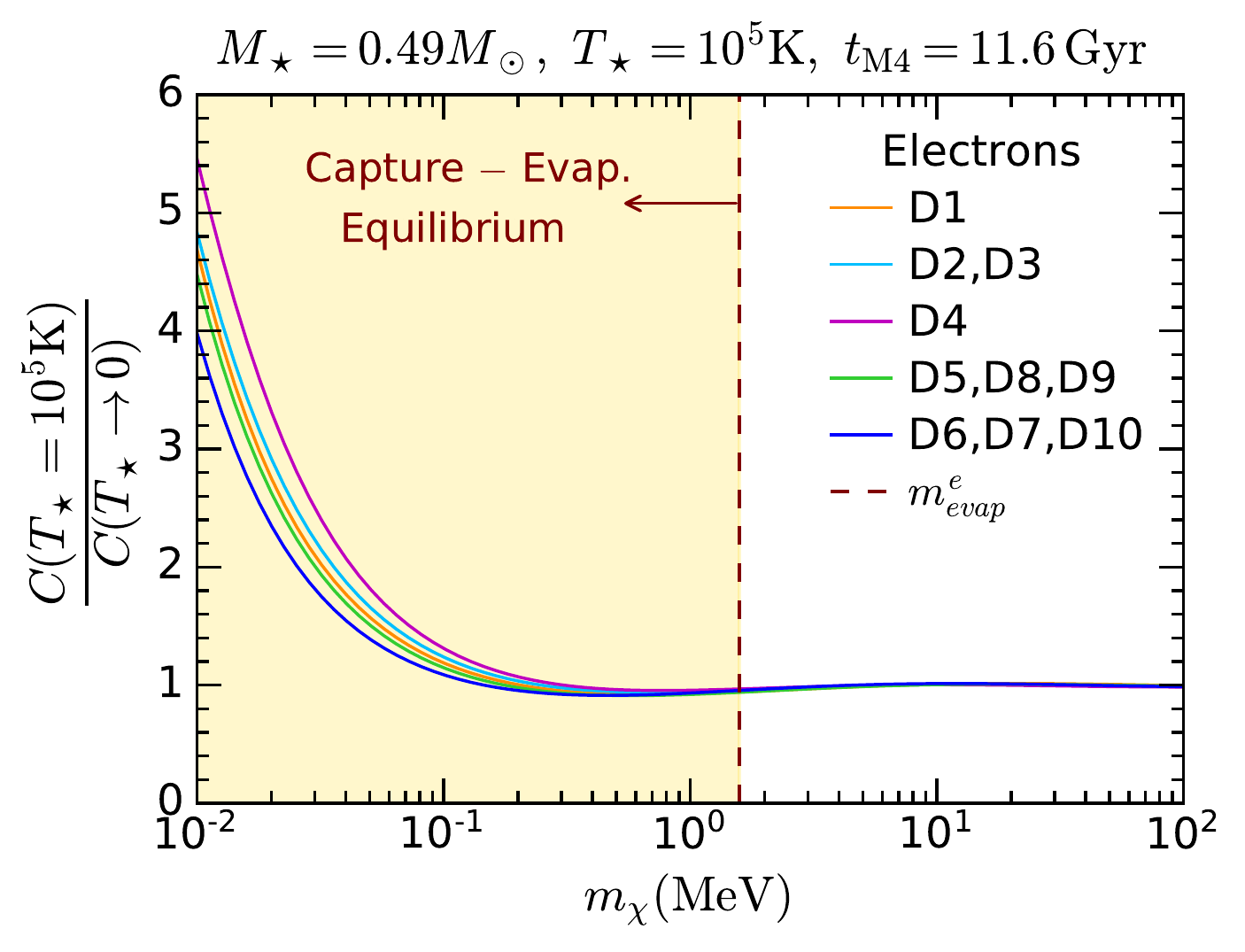}
  \includegraphics[width=0.495\textwidth]{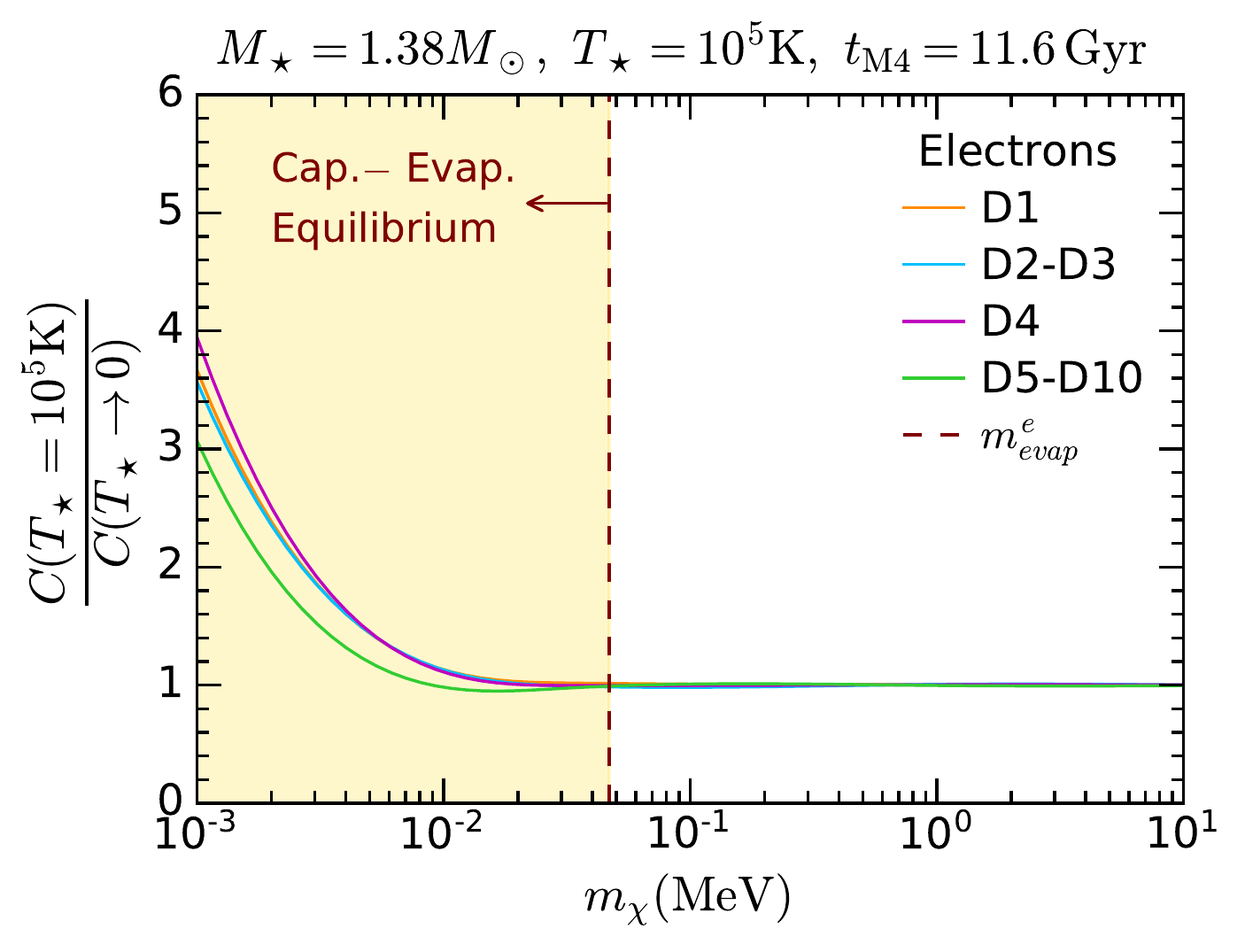} 
    \caption{Finite temperature effects on the capture rate in the case of scattering on electron targets, for two WDs in the globular cluster M4, namely WD$_1$ (left) and WD$_4$ (right) in Table~\ref{tab:WDs}. The DM mass range where capture and evaporation are expected to be in equilibrium is shaded in yellow. The dashed brown line corresponds to the evaporation mass.}
    \label{fig:capTfinelec}
\end{figure}

Accounting for the finite temperature of the star can have a large impact when the dark matter mass is low. There are two main effects. First, there is a boost in the capture rates as there are more available final states for the electrons to scatter into. 
Because the DM energy loss, $q_0$, is bounded from above to be $\qomax\lesssim3\MeV$ 
at zero temperature, only the outer shell of the Fermi sphere contributes to the capture rate. 
In comparison, non-zero $T_\star$ allows deeper shells of the Fermi sphere to contribute, which substantially increases the capture rate. 
Calculating the capture rate for finite $\Tstar$ is achieved by using the full form of the  Fermi-Dirac distributions in Eq.~\ref{eq:omegampauliur}, instead of approximating them with $\Theta$-functions, and removing the upper limit on the $E_e$ integration interval. 

Second, evaporation of the captured DM becomes possible due to scattering off the thermal electrons.  Again, this is relevant for low mass DM. 
In order for these finite temperature effects to be relevant in DM capture, they need to come into effect at DM masses above the evaporation mass of the WD~\cite{Garani:2018kkd, Bell:2020lmm}. 
To estimate the evaporation rate, we use  the full expression  obtained in ref.~\cite{Bell:2020lmm} for neutron stars, which can be approximated as 
\begin{equation}
E \sim  \frac{ m_\chi m_e^2\sigma_{e\chi}}{4\pi^2} \left(\frac{1}{\sqrt{B(0)}}-1\right)^2  \exp\left[{-\frac{m_\chi}{\Tstar}}\left(\frac{1}{\sqrt{B(0)}}-1\right)\right],  
\label{eq:evap}
\end{equation}
when the accreted DM is confined close to the centre of the star. 
We can then estimate the evaporation mass for scattering from electrons in the same manner as that outlined for scattering from ions in section~\ref{sec:evapions}.
Note that the evaporation rate is driven by the ratio $m_\chi/\Tstar$, and as such is enhanced for light DM. 

In Fig.~\ref{fig:capTfinelec}, we plot the ratio of the capture rate for $\Tstar=10^5\K$ to the zero temperature approximation for the same WDs as in Fig.~\ref{fig:Coptthinelec}. Operators that depend only on powers of the exchanged momentum $t$, namely D1-D4, are the most affected by finite temperature corrections, followed by  operators that contain in their matrix elements linear terms of the form $t^n s^m$, i.e., D5-D10. As can be immediately noticed, the DM mass range at which these effects become relevant depends on the specific WD configuration,  and is always below the evaporation mass for electron scattering $m_{evap}^e$ (dashed brown lines).  We find that the evaporation mass for carbon WDs with  $\Tstar=10^5\K$ ranges from $m_{evap}^e\sim 50\keV$ ($\Mstar=1.38\Msun$, right panel) to $m_{evap}^e\sim 1.5\MeV$ ($\Mstar=0.49\Msun$, left panel). 
As in the ion case, the evaporation mass is larger in warmer WDs, e.g. $\Tstar=10^6\K$, with an increase of one order of magnitude in $\Tstar$ resulting in a similar rise in $m_{evap}^e$.

\section{Results}
\label{sec:results}

In this section, we calculate bounds on the cutoff scale of the dimension 6 EFT operators that describe DM interactions with either electron or ion targets. To that end, we use the observed luminosity of the faintest WDs in the globular cluster M4~\cite{Bedin:2009,McCullough:2010ai}, together with the estimations for $\rho_\chi$, $\vstar$ and $v_d$ derived in ref.~\cite{McCullough:2010ai}.  We compute the capture rate in the optically thin limit, assuming that the WDs are made of $^{12}$C. Even though colder WDs have been recently observed by the Gaia mission~\cite{GentileFusillo:2019}, no significant bounds can be derived from this data due to the low DM density in the solar neighbourhood.

Once captured, the gravitationally bound DM will continue to scatter with the WD constituents until they reach thermal equilibrium. We have checked the thermalisation timescales for scattering on both electron and ion targets, using the method described in ref.~\cite{Bertoni:2013bsa}. For electrons, we find that the longest time to thermalise is $\sim 10^{5}$ yrs for the operators of interest. Scattering on ions requires additional input about the lattice structure to account for phonon emission/absorption processes~\cite{Acevedo:2019gre}. Though these effects have a significant effect on the thermalisation time, the timescales are still several orders of magnitude lower than the age of the WD (expected to be of the order of the age of M4).

Following thermalisation, the DM can self annihilate in the WD interior.  The number of DM particles present in the WD core therefore evolves as
\begin{equation}
    \frac{dN_\chi}{dt} = C - A N_\chi^2, 
    \label{eq:ndm}
\end{equation}
where the coefficient $A$ is related to the annihilation rate as
\begin{equation}
 \Gamma_{ann} =  \frac{1}{2}A N_\chi^2,   
\end{equation}
and we have assumed that evaporation is negligible, i.e., 
$m_\chi\geq m_{evap}$. 
The annihilation coefficients can be calculated from the thermally averaged annihilation cross sections for each operator, which can be found in ref.~\cite{Zheng:2010js}. To calculate these cross sections, as mentioned above we consider two separate cases, DM interactions with either quarks or leptons, characterised by the cutoff scale $\Lambda_q$ and $\Lambda_e$, respectively. This means that when computing 
bounds on $\Lambda_q$ for interactions with quarks, no assumptions were made regarding the strength of DM-lepton interactions. 
Instead, loop induced effective couplings to leptons 
were calculated in a similar fashion to refs.~\cite{Kopp:2009et,Bell:2019pyc}.  
Below the electron mass, annihilation to neutrinos, or loop induced annihilation to photons (non-zero only for some operators) are the only allowed annihilation channels. 

If the capture and annihilation processes are in equilibrium, then $\Gamma_{ann}=C/2$ and the DM contribution to the star luminosity is $L_\chi=m_\chi C(m_\chi,\Lambda_f)$. The time in which this equilibrium is reached is determined by the steady state solution of Eq.~\ref{eq:ndm}, and is given by
\begin{equation}
    \tau_{eq} = \frac{1}{\sqrt{C A}}.\label{eq:taueq}
\end{equation}
We can then set the EFT cutoff $\Lambda_f$ to the values obtained from the WD luminosity (see paragraph below)  to calculate the corresponding equilibrium times, and hence verify that capture-annihilation equilibrium is met. For electrons, the resulting times are significantly less than the age of the WDs. For ions, timescales longer than $1$ Myr are required to reach equilibrium in the case of operator D1 with DM mass $\lesssim 1\GeV$, while D2 can take as long as $10^4\yrs$ in the mass range of interest. The remaining operators all rapidly reach equilibrium. Given we are interested in old WDs, with ages of order $\sim$ Gyrs, we conclude that capture-annihilation is safely met for all cases of interest.

\begin{figure}
    \centering
    \includegraphics[width=\textwidth]{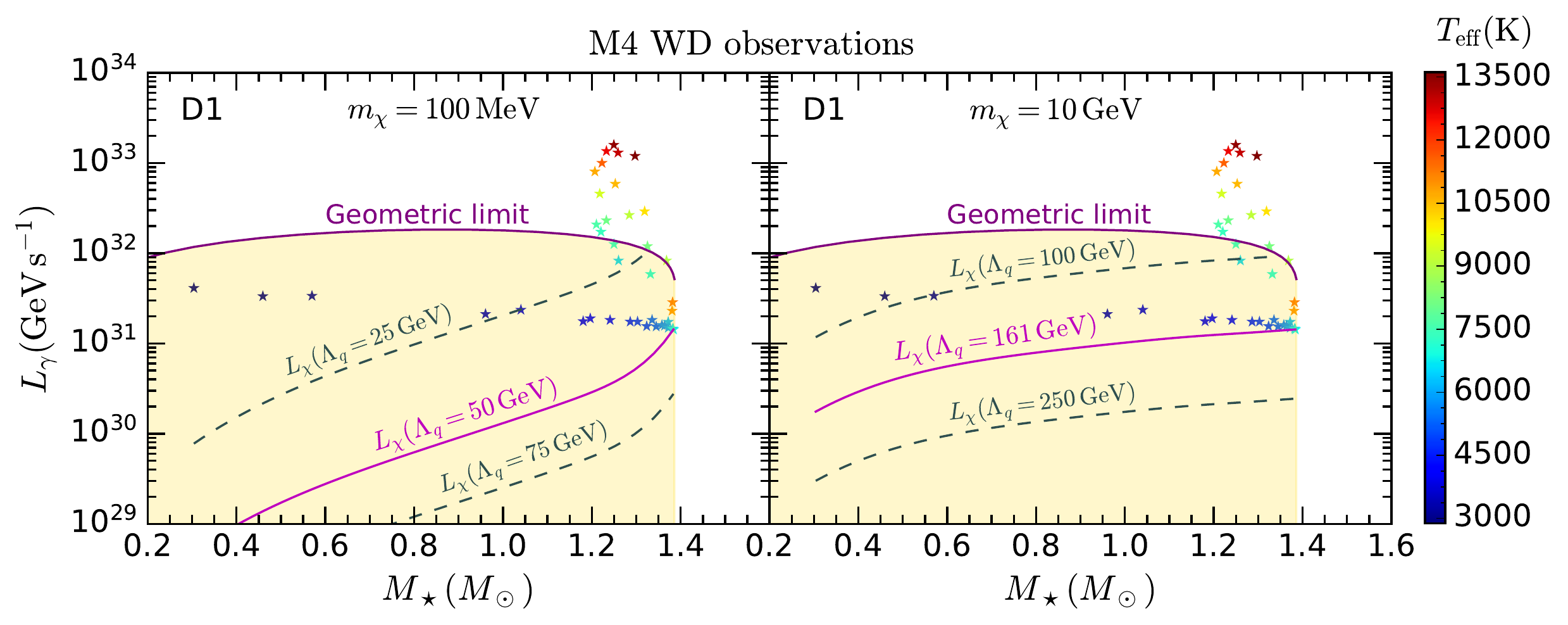}
    \caption{WDs observed in the globular cluster M4 and DM contribution to the star luminosity $L_\chi$ for different values of the cutoff scale $\Lambda_q$ and $m_\chi$ for D1. The dark violet lines correspond to the maximum achievable $L_\chi$ for nucleon targets, obtained in the geometric limit. The magenta lines represents the minimum value of $\Lambda_q$ that is consistent with the WD observations. }
    \label{fig:limitset}
\end{figure}

To estimate the limits on the cutoff scale $\Lambda_f$ for DM interactions with SM fermions, we compare the luminosity due to DM  with the WD observed luminosities $L_\gamma$. In Fig.~\ref{fig:limitset}, we illustrate how we have performed this calculation for DM scattering with nuclei, i.e., to determine $\Lambda_q$.  
The observed luminosity of the WDs in M4 is shown in the $L_\gamma-\Mstar$ plane\footnote{Actually, more WDs were observed in the globular cluster M4 than those shown in Fig.~\ref{fig:limitset}. We have given preference to the faintest WDs.}, where we have used the effective temperature to infer the radius of every star. Since, there are no independent measures of the mass of the WDs in M4 and we require radial profiles of the target number density, electron Fermi energy and escape velocity to compute capture and evaporation rates, we have 
solved the TOV equations coupled with the FMT EoS for carbon WDs to calculate $\Mstar$.\footnote{The mass and radius obtained using this method are in good agreement with recent observations within 2~kpc retrieved from the Montreal White Dwarf Database~\cite{Dufour:2017}, which contains more than 32000 WDs identified by Gaia DR2~\cite{GaiaDR2:2018} and EDR3~\cite{GaiaDR3:2020}, and spectroscopy measurements from surveys including SDSS DR12 and 2MASS.}  We also show the DM luminosity for different values of $\Lambda_q$ for $m_\chi=100\MeV$ (left) and $m_\chi=10\GeV$ (right), calculated using different WD configurations. 
As can be seen, the WD with $M_\star\simeq1.38 \Msun$ is the star that imposes a lower bound on $\Lambda_q$ (solid magenta line), since  $L_\gamma$ should be at least equal to the expected contribution from DM for all the observed WDs. 
In other words, if the luminosity due to DM capture and annihilation is at most equal to the observed luminosity of the faintest and heaviest WD in M4 ($M_\star\simeq1.38 \Msun$), there will be no tension between these observations and DM induced heating of WDs. 
While the results in Fig.~\ref{fig:limitset} assume WDs of a pure carbon composition, we have checked that a pure He composition for stars of $\Mstar\lesssim0.5\Msun$ does not alter the bounds on $\Lambda_q$. 
Note that the lower bounds are always well below the DM luminosity for maximal capture probability  (geometric limit, see purple lines). 
Lower values of $\Lambda_q$ (dashed grey lines) would be in tension with the lowest luminosity WDs.

\begin{figure}
    \centering
     \includegraphics[width=0.94\textwidth]{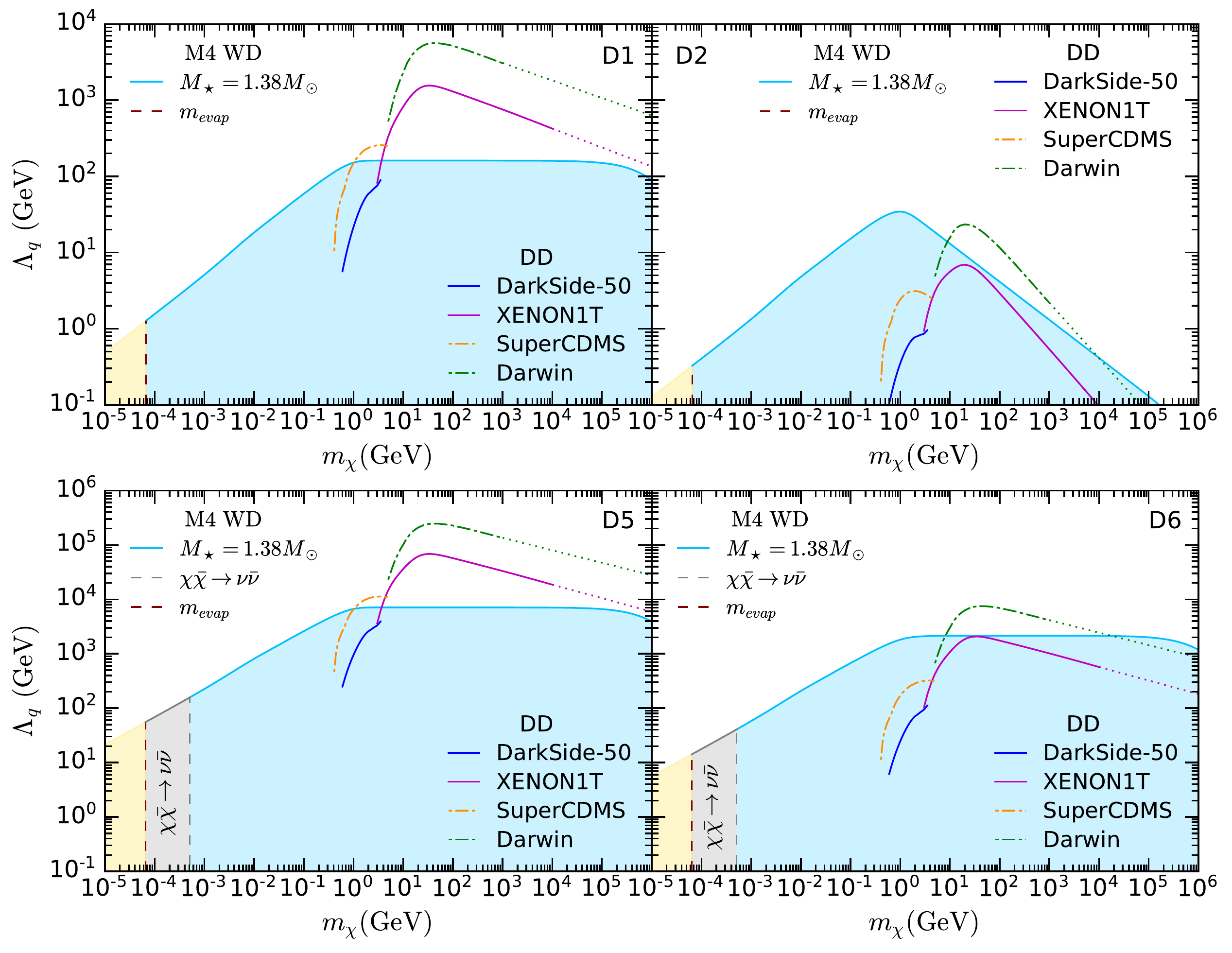}    
     \includegraphics[width=0.52\textwidth]{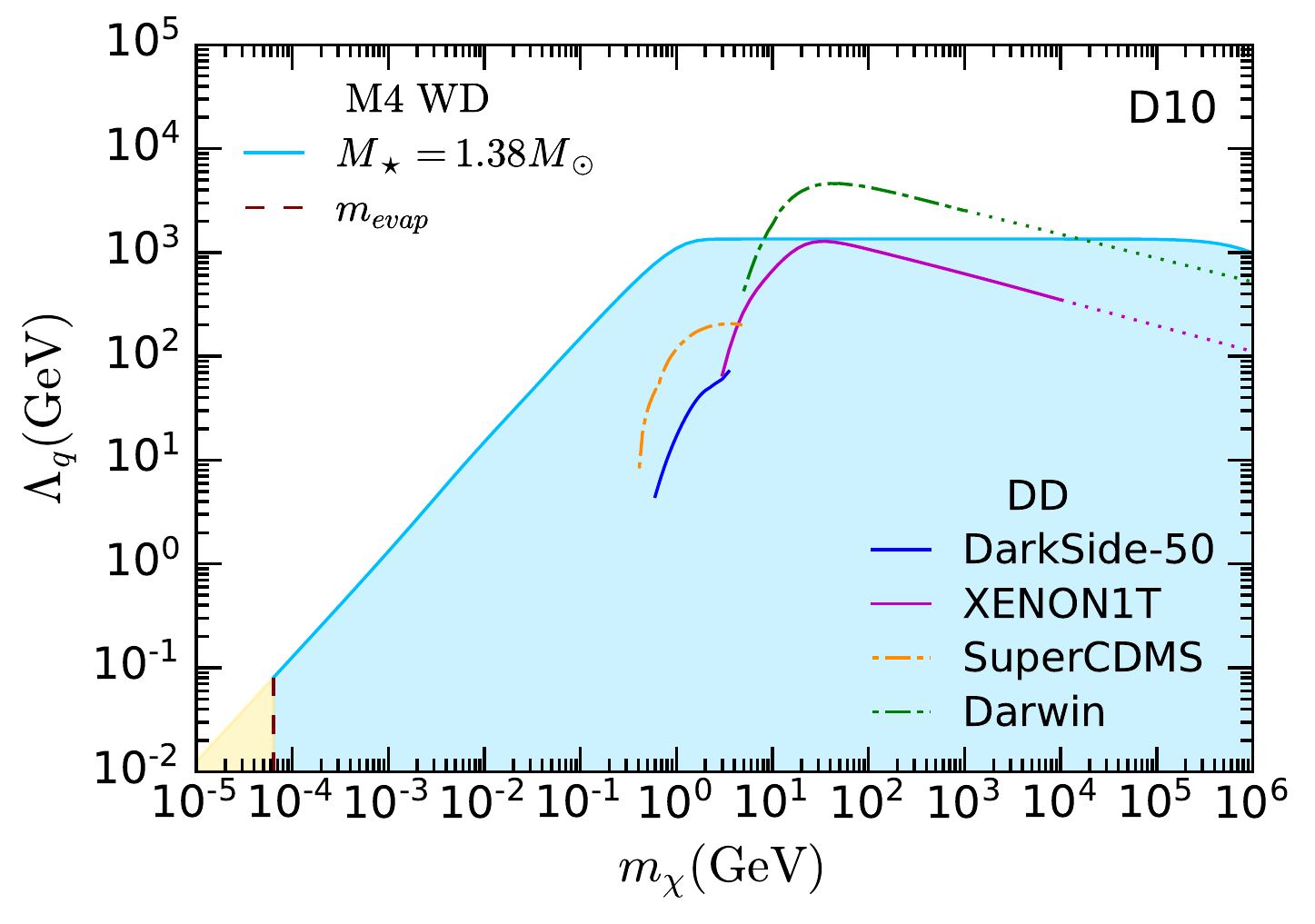}  
    \caption{Limits on $\Lambda_q$ for DM interactions with nucleons, obtained with the lowest luminosity and heaviest WD in the globular cluster M4, assuming $\rho_\chi=798\GeV\cm^{-3}$ (contracted halo) \cite{McCullough:2010ai} and $\Tstar=10^5\K$. The region where capture and evaporation are expected to be in equilibrium is shaded in yellow, and the region where DM annihilates to neutrinos that escape the WD is shaded in grey. 
    We also show lower bounds on $\Lambda_q$ from   DarkSide-50~\cite{Agnes:2018ves},  Xenon1T~\cite{Aprile:2018dbl,Aprile:2020thb}, and sensitivity projections from SuperCDMS~\cite{Agnese:2016cpb} and Darwin~\cite{Aalbers:2016jon}. 
    }
    \label{fig:LlimitsN}
\end{figure}

In Fig.~\ref{fig:LlimitsN}, we show the lower bounds on $\Lambda_q$ (light blue lines) for DM interacting with nuclei. As mentioned in section~\ref{sec:evapions}, we have assumed a core temperature $\Tstar=10^5\K$, which slightly affects the slope of the bounds at $m_\chi\lesssim10\MeV$ where the finite temperature effects come into play in the capture rate calculation.
The areas shaded in light blue represent the regions of the parameter space excluded by the observed luminosity of cold WDs in M4, provided that this globular cluster has been formed in a DM subhalo and the DM in the innermost region has survived tidal stripping as expected. 
These constraints are valid as long as $m_\chi \geq m_{evap}$,  where $m_{evap}$ has been calculated for the values of $\Lambda_q$ in Fig.~\ref{fig:LlimitsN} and $\tstar=t_{\rm M4}$.\footnote{Note that in  calculating the evaporation mass we have neglected DM annihilation; see section~\ref{sec:evapions}. However, we do not expect that a more refined approach 
will lead to a significant change in the bounds on $\Lambda_q$ for low mass DM, since the evaporation mass is determined mainly by the exponential term in Eq.\ref{eq:evap}.} Below this mass (yellow region) no limits on  $\Lambda_q$ can be derived using the WD luminosity. 
In addition, for operators D5 and D6 (middle panels), annihilation to neutrinos is the only channel open at $m_\chi\leq m_e$ (area shaded in grey). These neutrinos escape the WD without depositing energy in its interior and therefore we cannot derive limits from DM heating in this mass regime. 
For comparison, we also plot lower bounds  from the leading DM direct detection experiments, DarkSide-50~\cite{Agnes:2018ves} and  Xenon1T~\cite{Aprile:2018dbl,Aprile:2020thb}, and the projected sensitivity for Darwin~\cite{Aalbers:2016jon}. We can see that for the velocity and momentum suppressed operators (D2, D6 and D10), the WD limits surpass current bounds from direct detection. Only the future experiment Darwin can outperform the WD limits, and only in the region $10\GeV\lesssim m_\chi\lesssim1\TeV$. For the remaining operators, D1 and D5, Xenon1T limits are more stringent for $m_\chi\gtrsim5\GeV$. Note that the DM mass regime where WDs in M4 might outshine direct detection experiments, for all interaction types, is the low mass region starting at $m_\chi\lesssim5\GeV$ and extending down to the evaporation mass for operators D1, D2 and D10, and to the electron mass for D5 and D6.  For most of this parameter space, direct detection experiments have no sensitivity because the nuclear recoils would fall below experimental detection thresholds.

\begin{figure}
    \centering
  \includegraphics[width=0.7\textwidth]{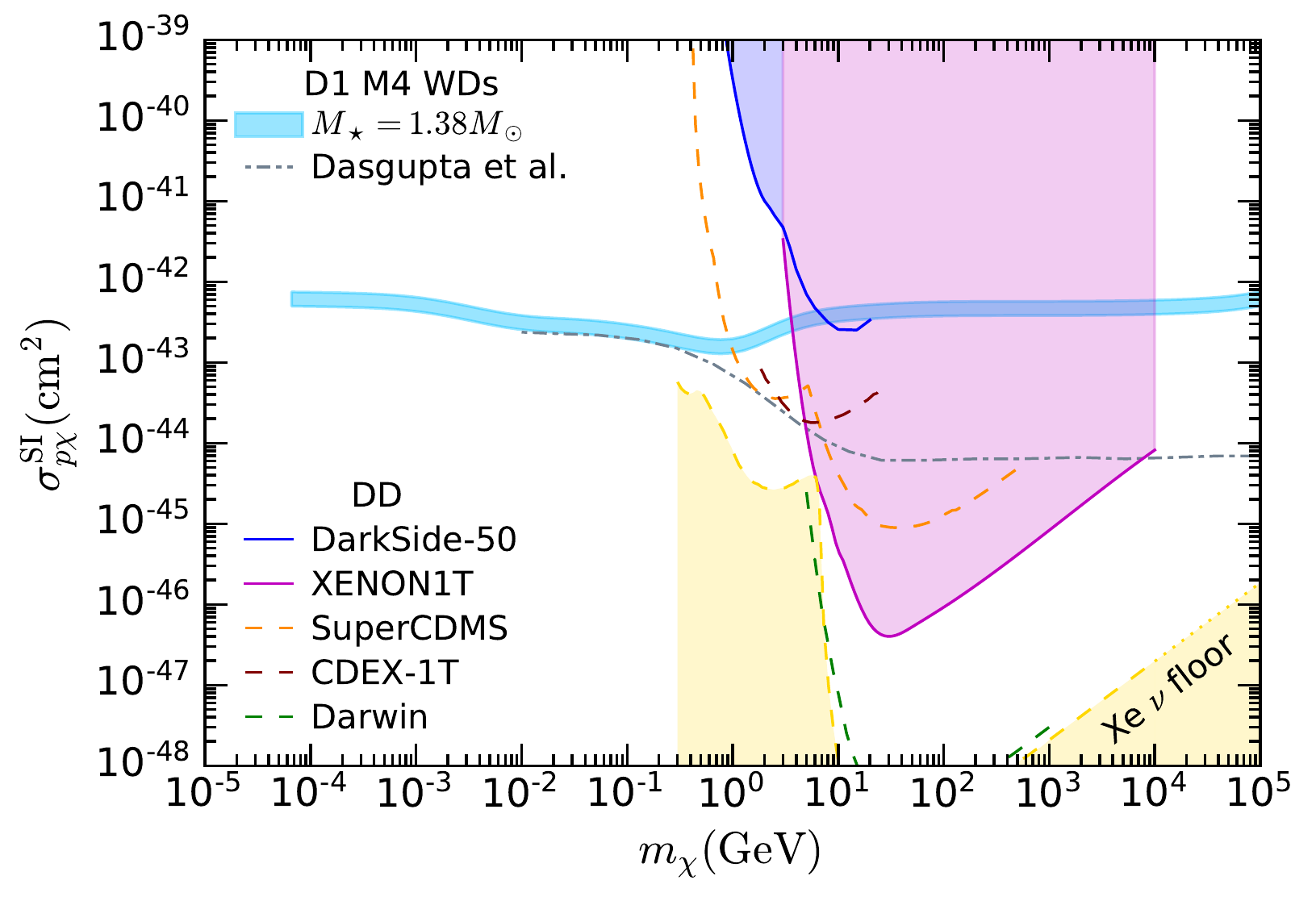}
    \caption{Spin independent DM-proton cross section for scalar-scalar interactions, $\sigma_{p\chi}^{\rm SI}$, as a function of the DM mass. The grey  line corresponds to the  limit on $\sigma_{p\chi}^{\rm SI}$ derived in ref.~\cite{Dasgupta:2019juq}, while the light blue band denotes the bound calculated using our approach for the capture rate computation in the optically thin limit.  The width of the blue band represents the uncertainty in the DM density in M4~\cite{McCullough:2010ai}.
    For comparison, we show direct detection limits from  DarkSide-50~\cite{Agnes:2018ves},  Xenon1T~\cite{Aprile:2018dbl,Aprile:2020thb},  projected sensitivities from SuperCDMS SNOLAB Ge/Si~\cite{Agnese:2016cpb}, CDEX-1T~\cite{Yue:2016epq}, and Darwin~\cite{Aalbers:2016jon}, as well as the neutrino coherent scattering background for xenon  detectors~\cite{Ruppin:2014bra}. 
    }
    \label{fig:D1xsecN}
\end{figure}

In Fig.~\ref{fig:D1xsecN}, we recast the lower limit on $\Lambda_q$ in terms of the DM-proton cross section for the scalar-scalar operator D1. The light blue band denotes the upper bound on $\sigma_{p\chi}$ considering the uncertainty in the DM density, which we take to range from $\rho_\chi\simeq532\GeV\cm^{-3}$ (uncontracted) to $\rho_\chi=798\GeV\cm^{-3}$ (contracted) for an NFW halo profile~\cite{McCullough:2010ai}. If the presence of DM in M4 is confirmed, this constraint will surpass those from direct detection, especially in the sub-GeV regime. 
The change of the slope in the band due to the WD core temperature is now evident at $m_\chi\sim10\MeV$. We also show the bound obtained in ref.~\cite{Dasgupta:2019juq} (dot-dashed grey line), using the same WDs in M4, though without taking into account either the WD internal structure or the nuclear response function for carbon. The difference in the shape of the bounds is due to the fact that the constraint in ref.~\cite{Dasgupta:2019juq} was derived from a light WD with $\Rstar\sim9000\km$, while we find instead that the WD with mass $\Mstar\simeq1.38\Msun$ provides the strongest upper bound on the DM-nucleon cross section (see Fig.~\ref{fig:limitset}).

\begin{figure}[t]
    \centering
    \includegraphics[width=0.98\textwidth]{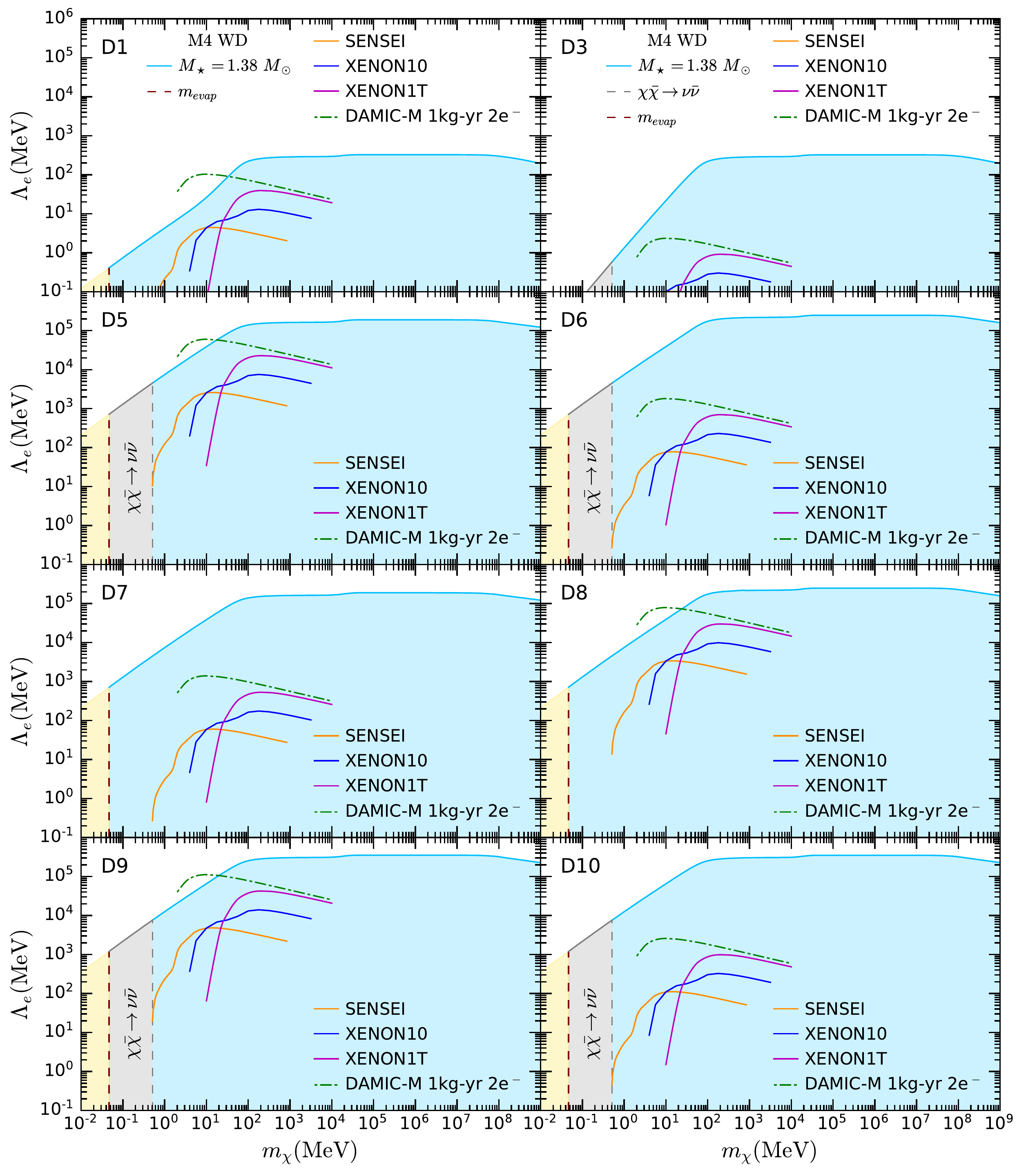}
    \caption{Limits on $\Lambda_e$ for DM interactions with electrons for the same WD as in Fig.~\ref{fig:LlimitsN}. The region where capture and evaporation (for $\Tstar=10^5\K$) are expected to be in equilibrium is shaded in yellow, and the region where DM annihilates to neutrinos that escape the WD is shaded in grey.
       For comparison, we show upper bounds from the leading electron recoil experiments for heavy mediators from SENSEI~\cite{Barak:2020fql}/DAMIC~\cite{Aguilar-Arevalo:2019wdi}, Xenon10~\cite{Essig:2017kqs}, Xenon1T~\cite{Aprile:2019xxb} and the projected sensitivity for DAMIC-M~\cite{Essig:2015cda}. }
    \label{fig:Llimitse}
\end{figure}

In Fig.~\ref{fig:Llimitse}, we show the limits on the cutoff scale $\Lambda_e$, in the case where DM is captured solely by collisions with the degenerate electrons. The shaded blue regions are the excluded parameters.
For operators D1-D4, for which the squared matrix elements
depend exclusively on the transferred momentum $t$, the DM-electron couplings is proportional to the tiny electron Yukawa coupling. 
This reduces the capture rate in such a way that for operator D4, the bounds on $\Lambda_e$ lie entirely in the $\Lambda_e \lesssim m_\chi$ region and for D2 only a small corner of the allowed parameter space surpasses this threshold. Given these low limits on the EFT cutoff scale $\Lambda_e$, such that an EFT description would not be valid, we do not plot results for D4 or D2.
For the remaining operators, especially D5-D10, there is a much larger region of parameter space where the limits on $\Lambda_e$ are such that an EFT description would be valid. In all cases, the lower limits on $\Lambda_e$, obtained using the WD with $\Mstar=1.38\Msun$ in M4, outperform the leading bounds from electron recoil experiments by at least $\sim$ an order of magnitude.  In most cases, they even outshine the projected sensitivity for the future experiment DAMIC-M (with the exceptions being D1, D5, D8 and D9 in the region below $m_\chi\lesssim10\MeV$ -- see dot-dashed green lines). DM scattering on electrons
is heavily hampered by Pauli blocking for  $m_\chi\lesssim100\MeV$.  The reach in the light DM mass regime is restricted by the evaporation mass for operators D1, D7 and D8 (see yellow region) and the electron mass for D3, D5, D6, D9 and D10 (see region shaded in grey) where DM annihilation to neutrinos is either the only final state allowed or the dominant channel.  Despite those limitations, we conclude that constraints from the observed luminosity of cold faint WDs in old globular clusters that have been able to retain their initial DM content in the innermost region of the cluster, can potentially exclude larger regions of the parameter space than direct detection, particularly in the sub-GeV region. This is especially relevant for leptophilic DM models.

\begin{figure}
    \centering
  \includegraphics[width=0.7\textwidth]{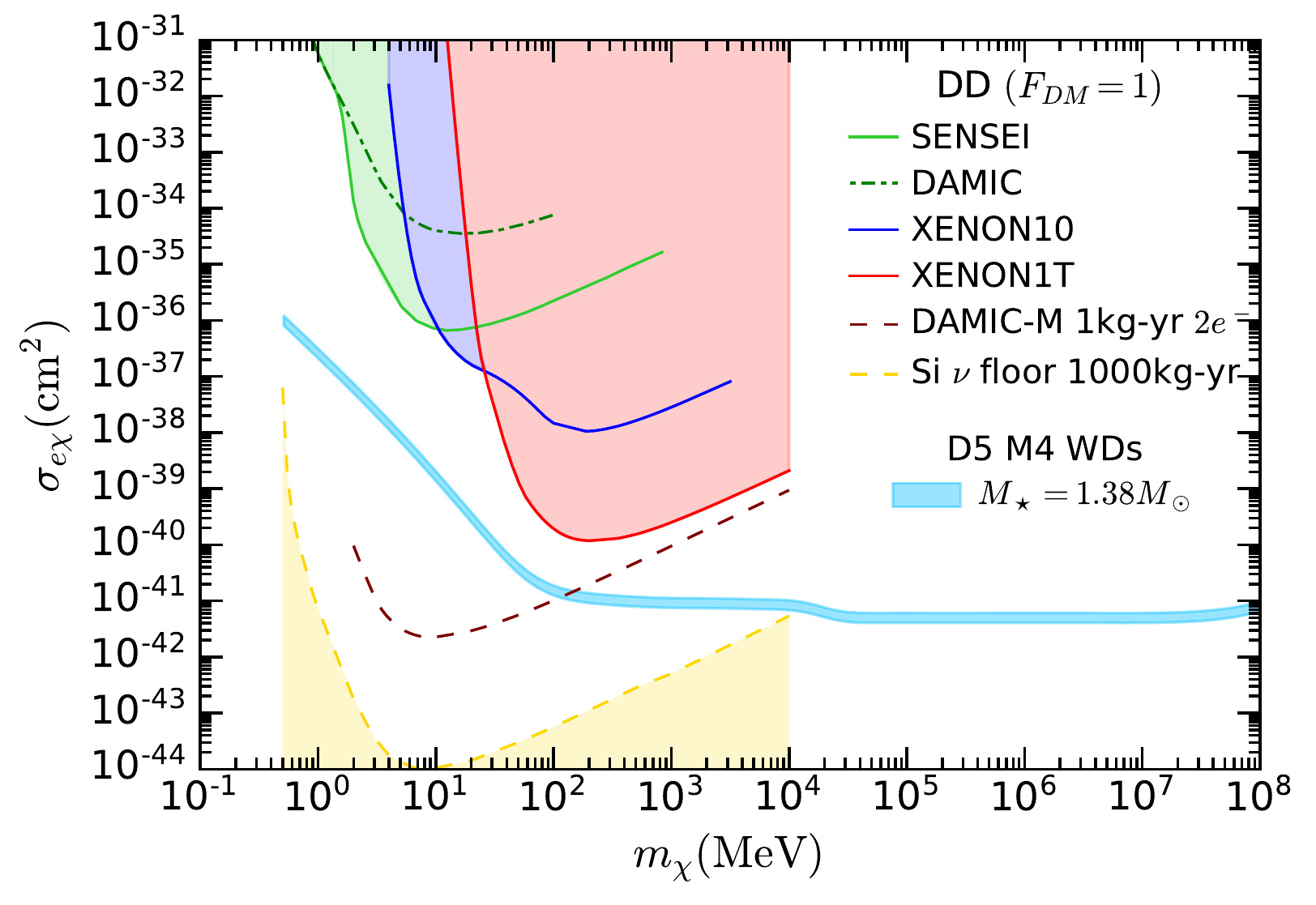}
    \caption{Upper bounds on the DM-electron scattering cross section for D5. The light blue band represents the limit from the WD observations in M4. The band width denotes the uncertainty in the DM density in M4~\cite{McCullough:2010ai}.  
     For comparison we show leading electron recoil bounds for heavy mediators from SENSEI~\cite{Barak:2020fql}, DAMIC~\cite{Aguilar-Arevalo:2019wdi}, Xenon10~\cite{Essig:2017kqs}, Xenon1T~\cite{Aprile:2019xxb}, the projected sensitivity for DAMIC-M~\cite{Essig:2015cda}, and the neutrino floor for silicon detectors~\cite{Essig:2018tss}. 
    }
    \label{fig:D5sigmalimit}
\end{figure}

Finally, in Fig.~\ref{fig:D5sigmalimit}, we conservatively compare the bound on the scattering cross section of the vector-vector operator obtained from WDs in M4, with the limits from electron recoil experiments. Even though the WD constraint is not able to probe the region where neutrino coherent scattering is expected to hamper the sensitivity of silicon detectors, or extend down below the electron mass, it would certainly surpass current DD bounds by orders of magnitude in $\sigma_{e\chi}$. It would even surpass  the projected sensitivity  for DAMIC-M, especially in the sub-MeV regime where no projections have been made\footnote{In the sub-MeV DM mass regime, modest limits on the DM-electron scattering cross section can be obtained by considering DM upscattered by cosmic rays. See, e.g. refs.~\cite{Cappiello:2018hsu,Ema:2018bih,Dent:2020syp}.}, despite the reduced WD sensitivity in this region due to Pauli blocking.

\section{Discussion and Conclusions}
\label{sec:conclusions}

White dwarfs (WDs) are the most abundant compact stars in the Galaxy. They are composed primarily of ions, usually carbon and oxygen nuclei, and supported against gravitational collapse by electron degeneracy pressure. 
Unlike neutron stars, their mass-radius relation is well-defined such that there is considerably less uncertainty in their equation of state (EoS). Moreover, old faint WDs have been observed in globular clusters and in the solar neighbourhood. In regions of high dark matter density, 
such as the inner Galaxy, 
DM capture and its subsequent annihilation in the stellar core can inject enough energy into these stellar remnants to prevent them from cooling. Therefore, anomalously bright isolated WDs observed in these regions may shed light on the nature of accreted DM and, conversely, the absence of such observations can be used to constrain DM models.  

In this paper, we improve previous calculations of the DM capture rate in WDs, for DM scattering either with nuclei in the stellar interior, or with the degenerate electron component.  For the case of scattering with nuclei, we include realistic nuclear response functions beyond the Helm approximation, incorporate the WD internal structure obtained with the relativistic Feynman-Metropolis-Teller EoS, and account for the star opacity. Unlike previous calculations, which were performed in the zero temperature limit, we evaluate the size of finite temperature corrections on the capture rate, which mainly affect the capture of light DM.  For DM capture due to collisions with relativistic degenerate electrons, we adapt the relativistic framework developed for neutron stars, which properly accounts for effects such as Pauli blocking, multiple scattering and general relativistic corrections.

We consider the nearest globular cluster, M4, where old, cold WDs have been observed by the Hubble Space Telescope. Assuming there is DM in the innermost region of M4, as suggested by simulations, we derive conservative upper bounds on the scattering cross sections for fermionic DM interactions with regular matter, or, equivalently, lower bounds on the cutoff scale of the effective operators that describe those interactions.

We find that old massive WDs in M4 can set constraints on DM-nucleon spin-independent cross sections that surpass present and future direct detection (DD) experiments.  This is especially so in the sub-GeV mass regime, or for velocity and momentum suppressed operators, where the WD bounds excel. This advantage of WDs over direct detection experiments is due to the lighter targets (carbon nuclei), the larger momentum transfers powered by the stronger gravitational pull of WDs, and the absence of a lower limit set by a recoil detection threshold.  Indeed, the WD limits extend down to the sub-MeV DM mass regime, with a lower cut-off that is determined by either the evaporation mass or the electron mass, depending on the interaction.

Similarly, when considering DM interactions with degenerate electrons, the observed luminosity of WDs in M4 leads to stronger bounds than the leading electron recoil experiments, despite DM capture by electrons being strongly restricted  by Pauli blocking in the $m_\chi\lesssim100\MeV$ region. These results are of particular importance for leptophilic DM, for which couplings to nucleons are loop suppressed.

We provide, for the first time, estimates of the evaporation rate and evaporation mass, for scattering on both ions and electrons. The evaporation mass is defined as the DM mass below  which capture and evaporation processes are expected to reach equilibrium. Bounds from WDs in the sub-MeV mass regime are restricted by the evaporation mass which, for old heavy WDs, is $\mathcal{O}(0.1)$~MeV for both DM-nucleon and DM-electron scattering. 

Finally, further simulations of galaxy formation and evolution, together with observations, will enable the existence of DM in M4 to be validated and improve the limits estimated here. Even if the presence of DM in the M4 globular cluster were ruled out, we have shown that, in general, WDs located in regions of high DM density are excellent probes of DM scattering due to their strong gravitational field.  Their reach is surpassed only by that of neutron stars, which are significantly harder to detect. 

\bigskip
\medskip

\noindent{\it Note added: Ref.~\citep{Garani:2021feo} appeared during the final stages of preparation of our manuscript. This paper deals with evaporation in a wide variety of stellar and substellar objects, including white dwarfs, but only for DM interactions with nuclei. Our calculation of the evaporation rate differs from ref.~\citep{Garani:2021feo}  mainly in the treatment of the stellar structure and DM-ion cross sections. 
}

\section*{Acknowledgements}
We thank Tony Thomas for helpful discussions. 
NFB and SR were supported by the Australian Research Council through the ARC Centre of Excellence for Dark Matter Particle Physics, CE200100008.
MV was supported by an Australian Government Research Training Program Scholarship and MRQ was supported by Consejo Nacional de Ciencia y Tecnologia, Mexico (CONACyT) under grant 440771.

\appendix
\section{Kinematic phase space for DM-electron scattering}
\label{sec:phasespace}

When deriving the interaction rate in the context of DM capture in NSs, we assumed that all the target phase space was available to scatter with DM. This is not necessarily true if the target is highly degenerate or if gravity is not particularly strong, like in NSs. In fact, for the scattering to occur, both the DM and target momenta should be in the inbound direction in the centre of mass frame. This is true in the whole phase space only when the 
following condition is satisfied 
\begin{eqnarray}
\frac{1}{\sqrt{B(r)}}>\frac{\muFe}{m_e}.
\end{eqnarray}

To clarify this point, we first derive this constraint using non-relativistic kinematics.
The DM particle of mass $m_\chi$ has an initial speed
\begin{equation}
    v_{esc}=\sqrt{1-B}, 
\end{equation}
while the target has a mass $m_e$ and an energy
\begin{equation}
    E_e= m_e+b\muFe=m_e\left(1+b \frac{\muFe}{m_e}\right),\quad b\in[0,1].
\end{equation}
Using non-relativistic kinematics, the speed of the target is 
\begin{equation}
    v_e=\sqrt{2b\frac{\muFe}{m_e}}.
\end{equation} 
The centre of mass velocity is defined as
\begin{equation}
\vec{v}_{com} = \frac{1}{m_\chi+m_e} (m_\chi \vec{v}_{esc}+m_e \vec{v}_e), 
\end{equation}
so the DM speed in the CoM frame is $ \vec{v}_\chi^{'} = \vec{v}_{esc}-\vec{v}_{com}$. 
To ensure that the DM and the target are not moving away from each other, the following condition should hold, $\vec{v}_\chi^{'}\cdot \vec{v}_{esc} >0$, 
i.e., the component of the DM velocity in the CoM frame is always parallel to the initial speed in the star frame. 
This condition leads to
\begin{equation}
    \cos\theta < \sqrt{\frac{1-B}{2b (\muFe/m_e)}}, \label{eq:condnonrel}
\end{equation}
where $\theta$ is the angle between the DM and the target speed in the star frame. If $\muFe/m_e\gg1-B$, the condition reduces to 
\begin{equation}
    \cos\theta<0, 
\end{equation}
meaning that the collision is head-on only. 

Repeating the same exercise with relativistic kinematics, 
the variable whose parameter space is modified by the above mentioned condition is the centre of mass energy $s$,
\begin{equation}
s = m_\chi^2+m_e^2+\frac{2m_em_\chi}{\sqrt{B}}\left(1+b\frac{\muFe}{m_e}-a\sqrt{1-B}\sqrt{2b\frac{\muFe}{m_e}+b^2\frac{\muFe^2}{m_e^2}}\right), \,  a\in[-1,1], 
\end{equation}
and the new condition becomes
\begin{equation}
a= \frac{s_{max}+s_{min}-2s}{s_{max}-s_{min}} <\left(1+b\frac{\muFe}{m_e}\right)\sqrt{\frac{1-B}{2b\frac{\muFe}{m_e}+b^2\frac{\muFe^2}{m_e^2}}}
= E_e\sqrt{\frac{1-B}{E_e^2-m_e^2}}.\label{eq:condrel}
\end{equation}

With the exception of the heaviest WDs, we have $\muFe/m_e\lesssim 1$. Then, with $1-B\sim10^{-3}$, we can check that expanding Eq.~\ref{eq:condrel} leads to Eq.~\ref{eq:condnonrel} and 
\begin{equation}
   a\lesssim 0.
\end{equation}
This implies that approximately half of the phase space is not available for scattering. (For electrons in NSs we instead have $\muFe/m_e\gg1$, and hence this does not occur.) Taking the ultra-relativistic limit for electrons, Eq.~\ref{eq:condrel} reduces to
\begin{equation}
a <\sqrt{1-B}\label{eq:condurel}.
\end{equation}
This restriction of the available phase space results in variations of the order of $\mathcal{O}(10\%)$ for both the interaction and capture rates. 

\section{Nucleon couplings for scattering operators}
\label{sec:operators}

Ten dimension six effective operators for fermionic DM interacting with quarks can be constructed, without considering flavour violation (see Table~\ref{tab:operatorshe}). The coefficients for the squared matrix elements in the fourth column of Table~\ref{tab:operatorshe}  read, 
\begin{eqnarray}
c_N^S &=& \frac{\sqrt{2}m_N}{v}\left[\sum_{q=u,d,s}f_{T_q}^{(N)}+\frac{2}{9}f_{T_G}^{(N)}\right],\\
c_N^P &=& \frac{\sqrt{2}m_N}{v}\left[\sum_{q=u,d,s}\left(1-3\frac{\overline{m}}{m_q}\right)\Delta_q^{(N)}\right],\\
c_N^V &=& 3,\\
c_N^A &=& \sum_{q=u,d,s}\Delta_q^{(N)},\\
c_N^T &=& \sum_{q=u,d,s}\delta_q^{(N)},
\end{eqnarray}
where $N=p,n$, $v=246$ GeV is the EW vacuum expectation value, $\overline{m}\equiv(1/m_u+1/m_d+1/m_s)^{-1}$ and $f_{T_q}^{(N)}$, $f_{T_G}^{(N)}$, $\Delta_q^{(N)}$ and $\delta_q^{(N)}$ are the hadronic matrix elements, determined either experimentally or by lattice QCD simulations. 
The values of the hadronic matrix elements for neutrons and protons used in this paper are listed in Table~\ref{tab:hadmatelem}.

\begin{table}[tb]
    \centering
    \begin{tabular}{|c|c|c|c|c|}
    \hline
     $q$ & $f_{T_q}^{(n)}$ \cite{Belanger:2013oya} & $f_{T_q}^{(p)}$ \cite{Belanger:2013oya} & $\Delta^{(n)}_q$ & $\delta^{(n)}_q$ \cite{Belanger:2013oya} \\ 
     \hline
     $u$  & 0.0110 & 0.0153     & -0.319 \cite{QCDSF:2011aa} & -0.230 \\ \hline
     $d$ & 0.0273  & 0.0191     & 0.787 \cite{QCDSF:2011aa} & 0.840 \\ \hline
     $s$ & 0.0447  & 0.0447     & -0.040 \cite{Dienes:2013xya} & -0.046 \\ \hline
    \end{tabular}
    \caption{Hadronic matrix elements for neutrons and protons. 
    The $\Delta^{(p)}_q$ can be obtained through $\Delta^{(p)}_{u,d} = \Delta^{(n)}_{d,u}$, and $\Delta^{(p)}_s = \Delta^{(n)}_s$ (similarly for the $\delta^{(p)}_q$). Also note that $f^{(N)}_{T_G} = 1 - \sum_q f^{(N)}_{T_q} = 0.917$.
    }
    \label{tab:hadmatelem}
\end{table}


\label{Bibliography}

\lhead{\emph{Bibliography}} 

\bibliography{Bibliography} 

\end{document}